\documentclass[iop]{emulateapj}

\shorttitle{HST Measures of Macc in the ONC}
\shortauthors{Manara et al.}

\def\macc   {$\dot{M}_{\rm acc}$}
\def\lacc   {$L_{\rm acc}$}

\begin{document}

\submitted{Submitted on May, 3$^{rd}$ 2012 - Accepted on June, 19$^{th}$ 2012}

\title{HST Measures of Mass Accretion Rates in the Orion Nebula Cluster}

\author{C.~F.~Manara\altaffilmark{1,2,4}, M.~Robberto\altaffilmark{1}, N.~Da Rio\altaffilmark{3}, G.~Lodato\altaffilmark{4}, L.~A.~Hillenbrand\altaffilmark{5}, K.~G.~Stassun\altaffilmark{6,7,8}, D.~R.~Soderblom\altaffilmark{1}}
\affil{$^1$Space Telescope Science Institute, 3700 San Martin Dr., Baltimore MD, 21218, USA\\
$^2$ European Southern Observatory, K. Schwarzschild-Str. 2, 85748 Garching, Germany \\
 $^3$European Space Agency, Keplerlaan 1, 2200 AG Noordwijk, The Netherlands\\
 $^4$Dipartimento di Fisica, Universit\'{a} Degli Studi di Milano, Via Celoria, 16, Milano, 20133, Italy\\
 $^5$California Institute of Technology, 1200 East California Boulervard, 91125 Pasadena, CA, USA\\
 $^6$Vanderbilt Univ., Dept. of Physics \& Astronomy 6301 Stevenson Center Ln., Nashville, TN 37235, USA}

\email{For comments: cmanara@eso.org}
\altaffiltext{7}{Fisk University, Department of Physics, 1000 17th Ave. N., Nashville, TN 37208, USA}
\altaffiltext{8}{Massachusetts Institute of Technology, Department of Physics, 77 Massachusetts Ave., Cambridge, MA 02139, USA}

\begin{abstract}
The present observational understanding of the evolution of the mass accretion rates (\macc) in pre-main sequence stars is limited by the lack of accurate measurements of \macc\ over homogeneous and large statistical samples of young stars. Such observational effort is needed to properly constrain the theory of star formation and disk evolution. Based on HST/WFPC2 observations, we present a study of \macc\ for a sample of $\sim$ 700 sources in the Orion Nebula Cluster, ranging from the Hydrogen-burning limit to $M_*\sim 2 M_\odot$. We derive \macc \ from both the $U$-band excess and the $H\alpha$ luminosity ($L_{H\alpha}$), after determining empirically both the shape of the typical accretion spectrum across the Balmer jump and the relation between the accretion luminosity ($L_{\rm acc}$) and $L_{H\alpha}$, that is $L_{\rm acc}/L_\odot = (1.31 \pm 0.03) \cdot L_{H\alpha}/L_\odot$ + (2.63$\pm$ 0.13). Given our large statistical sample, we are able to accurately investigate relations between \macc\ and the parameters of the central star such as mass and age. We clearly find \macc\ to increase with stellar mass, and decrease over evolutionary time, but we also find strong evidence that the decay of \macc\ with stellar age occurs over longer timescales for more massive PMS stars. Our best fit relation between these parameters is given by: $\log$(\macc/$M_\odot\cdot$yr)=(-5.12 $\pm$ 0.86) -(0.46 $\pm$ 0.13) $\cdot \log (t$/yr) -(5.75 $\pm$ 1.47)$\cdot \log (M_*/M_\odot)$ + (1.17 $\pm$ 0.23)$\cdot \log (t/$yr$) \cdot \log (M_*/M_\odot)$. These results also suggest that the similarity solution model could be revised for sources with $M_* \gtrsim 0.5 M_\odot$. Finally, we do not find a clear trend indicating environmental effects on the accretion properties of the sources.
\end{abstract}

\section{Introduction}
During the Pre-Main-Sequence (PMS) phase of stellar evolution, the interaction between the forming star and the surrounding disk is regulated by the accretion of disk material along the field lines of the stellar magnetosphere \citep{H98}. The gravitational energy released as the material falls along accretion columns and hits the stellar surface creates a characteristic shock spectrum \citep{Calvet}, with excess emission especially strong in the Balmer continuum and recombination lines \citep{Gullbring, Calvet}. The relative accretion luminosity (\lacc)  can be measured with spectroscopic \citep{Valenti93,Herczeg08} or photometric  \citep{Gullbring,Robberto,DeMarchi} methods. The mass accretion rate, \macc, is then estimated through the relation \citep{H98}:
\begin{equation}
\dot{M}_{\rm acc} = \frac{L_{\rm acc} R_*}{0.8 G M_*},
\label{Macc_eq}
\end{equation}
where $R_*$ and $M_*$ are the radius and the mass of the star, respectively, and the factor $0.8$ accounts for the assumption that the infall originates at a magnetospheric radius $R_{\rm m} = $5 \citep{Shu94}.

The mass accretion rate generally decreases over time during the first few Myr of PMS stellar evolution, as the circumstellar disks disperse their gaseous content on a timescale of $\sim$ 3-5 Myr \citep{Haisch01,Dahm05,Fedele}. Moreover, \macc\ is expected to scale with stellar mass.
The evolution of \macc \ vs. time as a function of the stellar and disk mass represents a key aspect of PMS evolution and planet formation.
\citet{H98} showed that while typical \macc \ values, at any given stellar age, can differ star by star of up to about two orders of magnitudes, on average they decrease exponentially with the stellar age $t$, i.e. \macc $\propto t^{-\eta}$ with $\eta \sim 1.5$. Several studies, targeting different young stellar clusters, have  confirmed this general trend (e.g. \citealt{Robberto}, \citealt{Aurora10}), but the uncertainties in the age estimate \citep{Hartm97,Baraffe10} and the scarcity of rich and homogeneous samples limit the accurate assessment of this dependence.

For what concerns the stellar mass dependence of \macc, \citet{Muzerolle03} found that $\dot{M}_{\rm acc}\propto M_*^b$ with $b=2$, although recent studies \citep[e.g.,][]{Rigliaco} suggest smaller values of $b$.
Due to the large spread in the \macc\ values for a given $M_*$ (up to two orders of magnitudes, \citealt{Rigliaco}), this second relation has also never been accurately constrained.
Although observational uncertainties and intrinsic variability will always contribute in scattering the measured mass accretion rates, accurate measurements of \macc \ on a large sample of PMS stars may allow us to assess more precisely the value of the two power law exponents ($\eta$ and $b$). This is a critical step, as they can be tied to the  theoretical model for the disk evolution and structure, providing unique constraints to the initial conditions of planet formation (e.g. \citealt{Alexander06}, \citealt{Lodato}).

As the nearest ($d\sim 414 \pm 7$ pc, \citealt{distanzaONC}) site of massive star formation, the Orion Nebula Cluster (ONC) provides a standard benchmark for star formation studies. The ONC population has been studied in depth, and determination of the individual stellar parameters are available for a large fraction of members. In particular  \citet{H97}, \citet{DaRio} and \citet{DaRio12} have used both spectroscopic and photometric techniques to derive the spectral type of more than 1700 ONC sources. The corresponding initial mass function, ranging from the Brown Dwarfs (BD) regime to the O6 star $\theta^1$Ori-C,  peaks at 0.2-0.3 $M_{\odot}$ \citep{DaRio12}, while the cluster mean age is $\sim$ 2.2 Myr, with evidence for an age spread of the order of $\sim$ 2 Myr \citep{Maddi}. Given the wealth of information on the individual cluster members, the ONC is ideally suited to conduct an extensive study of the mass accretion process in PMS stars.

In this paper we present the results of such a study based on the HST/WFPC2 survey of the Orion Nebula Cluster (GO 10246, P.I. M. Robberto). Both U-band and $H\alpha$ data are used to estimate \macc \ for $\sim$ 700 sources and to analyze the relations between this parameter and the main stellar parameters (age, $M_*$). In Sec.~\ref{observations_sec} we illustrate the observation, data reduction and analysis of our WFPC2 photometry. Sec.~\ref{method} presents our derived modeling of the photospheric colors for our sources and for the typical accretion spectrum, and the methods used to obtain the \lacc \ based on our data, both using our UBI Diagram method and from the $H\alpha$ photometry, while in \ref{results} we present the derived quantities (\lacc, \macc) obtained both from the measurement of the $U$-band or from $L_{H\alpha}$; in Sec.~\ref{analysis} we study the evolution of these parameters as a function the main stellar parameter. Finally, in Sec.~\ref{conclusion} we summarize our conclusions.

%________________________________________________
\section{Observations and data reduction}
\label{observations_sec}

%________________________
\subsection{Observations}
\label{observations}

In this work we consider the Hubble Space Telescope (HST) observations obtained for the HST Treasury Program on the ONC \citep{HST_treas}. In particular, we use the WFPC2 observations carried out between October 2004 and April 2005 in the filters F336W, F439W, F656N and F814W. One of the advantages of this data set is the nealry simultaneous (within less than one hour) imaging obtained in subsequent exposures in $U$ (F336W) and $H\alpha$ (F656N) bands, which, as will be described in the next sections, both allow an estimate of the accretion luminosity. 

\citet{HST_treas} details the data acquisition and reduction strategy of all the photometry obtained within the Treasury Program; here we summarize only the key aspects of the WFPC2 dataset, which are most relevant to our study.
The F336W filter is roughly analogous to the standard Str$\ddot{\rm o}$mgren \emph{u} filter, with a long wavelength cutoff at $\lambda \simeq$ 3600 \AA, shortward the Balmer jump. This WFPC2 filter, however, is affected by ``red leak'', which is a residual transmission at longer wavelengths ($\lambda \simeq$ 7300 \AA). This effect alters the measured $U-$band fluxes for red sources, and in Appendix~\ref{red_leak} we explain how the correction for this leak is performed. We hereafter refer to the corrected $U$ band as $U$, using the notation $U_{noleak}$ only when needed, for clarity. The F439W filter closely matches the standard Johnson $B$ passband, while the F656N filter is a narrow-band filter centered on the H$\alpha$ line, narrow enough to exclude the adjacent [NII] doublet lines. The last filter, F814W, is similar to the Cousin I$c$ filter; it is fundamental for our study both to determine the red leak contamination of the $U$-band, and to derive the bolometric luminosity of the ONC members. The exposure times for each field are 800~s in F336W, 80~s in F439W, 400~s in F656N  and 10~s exposure in F814W. 

Aperture photometry was obtained by choosing aperture radii of $0\farcs1$ and $0\farcs5$, corresponding to 5 and 11 pixels on the WF1 (PC) and 2 and 5 pixels on WF2-4 chips. The sky annulus has always been taken between $1\farcs$ and $1\farcs5$, corresponding to 20 and 30 pixels with WF1 and 10 and 15 pixels with WF2-4.
At the time the observations were carried out, the WFPC2 instrument had already accumulated about 11 years of total radiation dose in space environment, and therefore was affected by non-negligible Charge Transfer Efficiency (CTE) losses. The brightness of the Orion Nebula background, especially in our four broad-band filters and at the center of the region, mitigates the problem but still we had to apply a CTE correction to the measured counts following the recipe of \citet{CTEcorr}. We estimated CTE correction errors using a Monte Carlo propagation of the errors terms in the equation of \citet{CTEcorr}.

The final photometric errors have been derived by adding in quadrature the errors associated to the measured counts, zero point, CTE correction, aperture correction to the infinity and, in the case of the $0\farcs1$ apertures, an extra aperture correction to the $0\farcs5$ radius. This last term is usually dominating, ensuring that the $0\farcs5$ photometry is generally more accurate and therefore preferred, except for the faintest sources detected in very few pixels.

%________________________
\subsection{Source selection}
\label{selection_sources}

Our WFPC2 photometric catalog includes 1643 sources in total with detection in at least one of the four bands. Among these, we have removed 27 known proplyds, identified by \citet{Ricci_atlas} from the ACS imaging, since the fluxes of these are potentially contaminated by circumstellar emission. We have also eliminated 144 close binary systems ($d \lesssim$ 3 pix $\sim$ 75-100 AU), resolved either in the ACS or the WFPC2 imaging, where the spectral type classification is more uncertain. For well separated visual binaries, we carefully inspected the individual sources to rule out possible matching errors between the photometry in different bands.

In the remaining sample, $U$-band photometry is available for 1021 stars ($\sim$ 60\% of the total), among these, 897 sources ($\sim$ 55\%) have been detected in all $U$, $B$ and $I$ bands. Finally, for 1342 sources ($\sim$ 80\%) $H\alpha$ photometry is available.\\
\hbox{}\\
%\hbox{}\\
%\hbox{}\\
\hbox{}

%________________________
\section{Method}
\label{method}
We follow the approach introduced by \citet{DaRio} to obtain simultaneously an estimate of extinction ($A_V$) and accretion ($L_{\rm acc}/L_{\rm tot}$, where $L_{\rm tot} = L_*+L_{\rm acc}$) for individual sources from the photometry in 3 optical bands together with the $T_{\rm eff}$ of the source. The procedure assumes that on a color-color diagram the displacement of the observed sources from the theoretical isochrone is due to a combination of extinction and accretion. First of all accretion excess causes a shift in the colors towards bluer values, then extinction moves the resulting color along the reddening direction. In particular, in a two-color diagram, accretion displaces the sources along curves that start from the isochrone and converge to a point defined by the colors of the pure accretion spectrum. If $T_{\rm eff}$ of a star is known, e.g., from spectroscopy, the combination of $A_V$ and $L_{\rm acc}/L_{\rm tot}$ necessary to reproduce the observed colors is unequivocally found.

In \citet{DaRio} a $BVI$ two-color diagram was used; here instead we construct the $U_{\rm noleak}-B$ versus $B-I$ two-color diagram. Since $U$-band is far more sensitive to accretion than $B$, this allows us to measure $L_{\rm acc}/L_{\rm tot}$ with a significantly higher accuracy than in \citet{DaRio}. We will refer from now on to this two-color diagram as the 2CD.

\subsection{The Two-Colors Diagram: Extinction and Accretion Luminosity}
\begin{figure}
\epsscale{1}
\plotone{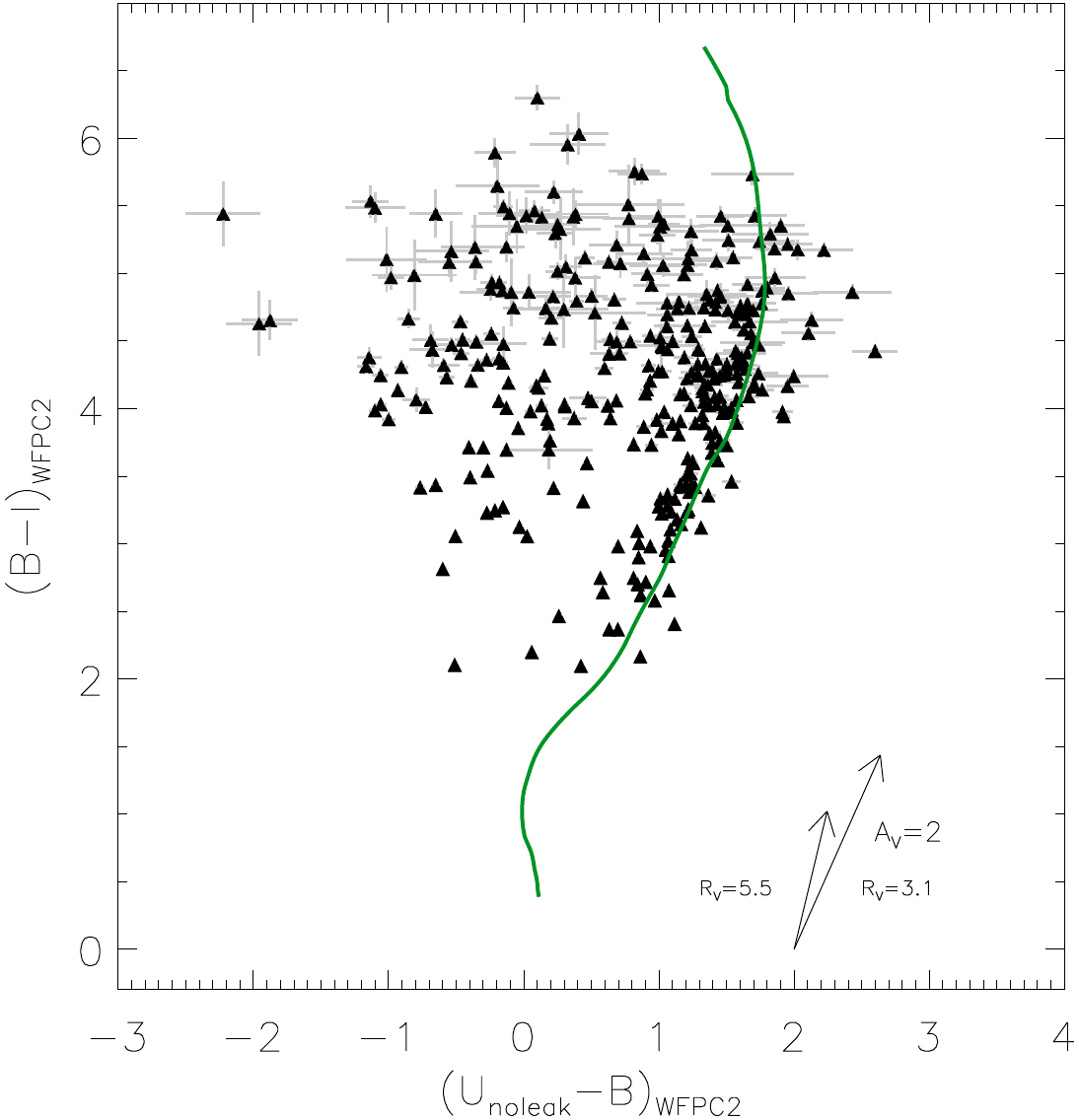}
\caption{Our $UBI$ 2CD; the thick line is our calibrated isochrone valid for the ONC. The arrows indicate two reddening vectors, corresponding to $A_V$=2, for two different values of $R_V$ assuming the extinction law of \citet{Cardelli}}
\label{UBI_ZAMS}
\end{figure}

We consider the measured $T_{\rm eff}$ for the ONC members from \citet{DaRio12}, and limit ourselves to the sources detected in all the three WFPC2 bands and where the error on the $U_{noleak}$ values is $\lesssim$0.3~mag\footnote{The photometrical errors in all the bands ($U,B, I)$, calculated as in Sec.~\ref{observations}, are always lower than 0.2~mag, but the correction for the red leak (see Appendix~\ref{red_leak}) raises the final error on the measure to values up to $\sim$0.29~mag in 15 objects.}. Our sample is then composed by 339 ONC members. Fig.~\ref{UBI_ZAMS} shows the observed 2CD of this sample; the solid line represents our calibrated isochrone (see Sec.~\ref{model_calibration} for details).

To model the extinction, we assume the reddening law of \citet{Cardelli} with typical galactic reddening parameter $R_V=3.1$. Some authors have suggested a larger value of $R_V=5.5$ for the ONC \citep[e.g.,][]{Rv5.5} and we show this as well in Fig.~\ref{UBI_ZAMS}. \citet{DaRio}, however, found $R_V=3.1$ to be more appropriate, on average, for the ONC population. Therefore, we will adopt this value for our analysis.

\subsubsection{Color Calibration}
\label{model_calibration}

As required by our method, we need to assume an isochrone that defines the intrinsic, photospheric, colors as a function of T$_{\rm eff}$, for the ONC sources in our 2CD. As shown by \citet{DaRio}, current atmosphere models are still unable to lead to accurate predictions of optical colors for young ($\sim$ 2 Myr) cool stars, and empirical calibrations of photospheric colors have not been specifically carried out for young PMS stars. Also, in that work the authors show that, for the same T$_{\rm eff}$, the photospheric colors of $\sim$ 2 Myr PMS sources are not in agreement with those of MS dwarfs. Thus we cannot either rely on calibrations of colors valid for more evolved populations. This problem therefore requires a particular care. 
We explain in Appendix~\ref{app_model_calibration} the procedure used to obtain the correct isochrone calibration for our 2CD, and we report in Table~\ref{corrections} the colors of the isochrone. It could be argued that the assumption of a single isochrone in our 2CD, given a possible age spread in Orion, may lead to false predictions. This is not the case, for 2 reasons: a) the 2CD is, by its nature, independent on stellar luminosities or radii; therefore age variations for the same T$_{\rm eff}$ can result in differences in the photospheric colors only due to changes in the stellar surface gravity; b) in the optical, the difference in colors between PMS and MS stars can be of up to several tenths of a magnitude \citep{DaRio}. This offset, scaled down to the modest age spread within the ONC \citep{Maddi}, results in negligible differences in the colors vs. T$_{\rm eff}$ relations.

In Fig.~\ref{UBI_ZAMS}, we present our calibrated isochrone, together with the observed sources. If ongoing mass accretion did not alter our observed colors, the sources would have been located on the model line, with only a spread given by extinction. Instead, a number of ONC members are displaced towards bluer colors, as expected by the presence of accretion flux excess.

For our subsequent analysis, and specifically for the derivation of bolometric stellar luminosities, we also require the bolometric corrections ($BC$) in the $I-$band as a function of $T_{\rm eff}$. These are defined as the difference between the bolometric magnitude of a star and the apparent $I-$band magnitude.
\begin{eqnarray}
BC = M_{bol} - M_{\lambda} = \nonumber \\
=  2.5 \log \left[\frac{\int f_{\lambda}(T)\cdot S_{\lambda} \ d\lambda}{\int f_{\lambda}(T) \cdot d\lambda}\middle/\frac{\int f_{\lambda}(\odot)\cdot S_{\lambda} \ d\lambda}{\int f_{\lambda}(\odot) \cdot d\lambda}\right].
\end{eqnarray}

We computed the BCs by performing synthetic photometry on the spectra of \citet{BT-Settle}, and corrected the results based on the same offsets we determined in our calibration of the colors, to correct the I-band intrinsic magnitudes.
The values of this empirical calibrated bolometric correction for various spectral types are also shown in Table \ref{corrections}.
The BC$_I$ varies between $\sim$ 0.03 and $\sim$ -0.70 magnitudes in the range of temperatures of our sources (3000 K $\textless \ T_{\rm eff} \ \textless$ 8000 K).

\begin{deluxetable}{ccccc}
\tablecaption{Intrinsic colors for the WFPC2 bands F336W (U), F439W (B) and F814W (I) and Bolometric Correction for the F814W (I) band obtained assuming~\citet{BT-Settle} models.}
\tablehead{\colhead{Spectral} & \colhead{$T_{\rm eff}$} & \colhead{($U-B$)} & \colhead{($B-I$)} & \colhead{($BC_{I}$)} \\
\colhead{Type} & \colhead{(K)} & \colhead{(mag)} & \colhead{(mag)} & \colhead{(mag)} }
\startdata
A5 & 8200 & 0.126 & 0.285 & -0.450 \\
A7 & 7850 & 0.106 & 0.424 & -0.365 \\
F0 & 7200 & 0.066 & 0.673 & -0.236 \\
F2 & 6890 & 0.017 & 0.809 & -0.180 \\
F5 & 6440 & -0.013 & 1.022 & -0.105 \\
F8 & 6200 & -0.002 & 1.151 & -0.068 \\
G0 & 6030 & 0.022 & 1.247 & -0.044 \\
G2 & 5860 & 0.054 & 1.352 & -0.021 \\
G5 & 5770 & 0.075 & 1.406 & -0.012 \\
G8 & 5570 & 0.142 & 1.530 & 0.007 \\
K1 & 5080 & 0.457 & 1.871 & 0.032 \\
K2 & 4900 & 0.597 & 2.024 & 0.033 \\
K3 & 4730 & 0.710 & 2.188 & 0.027 \\
K4 & 4590 & 0.790 & 2.342 & 0.017 \\
K5 & 4350 & 0.945 & 2.636 & -0.010 \\
K7 & 4060 & 1.111 & 2.997 & -0.055 \\
M0 & 3850 & 1.234 & 3.272 & -0.108 \\
M1 & 3705 & 1.313 & 3.460 & -0.159 \\
M2 & 3560 & 1.412 & 3.646 & -0.218 \\
M3 & 3415 & 1.543 & 3.890 & -0.288 \\
M4 & 3270 & 1.665 & 4.266 & -0.378 \\
M5 & 3125 & 1.771 & 4.722 & -0.515 \\
M6 & 2990 & 1.771 & 5.143 & -0.762 \\
M7 & 2880 & 1.734 & 5.513 & -1.050
\enddata
\label{corrections}
\end{deluxetable}

%-----------------------------------------
\subsubsection{Accretion spectrum}
Spectra of the flux excess produced by accretion have been modeled by  \citet{Calvet} (hereafter CG98) using a hydrodynamic treatment for the infalling material and heated photosphere.
It is possible, however, to reproduce a typical accretion spectrum using simpler recipes, for instance modelling the radiative transfer within a slab of dense gas \citep{Valenti93} or assuming suitable combination of an optically thin emission generated in the preshock region and an optically thick emission generated by the heated photosphere. In the latter case the two components can be added with a relative fraction of about 1/4 and 3/4, respectively \citep{Gullbring2000}. For simplicity, we follow this latter approach; for the optically thick component, we consider a 8000~K black body, and for the optically thin one, we modeled the emerging spectrum from a slab of gas using the version 10.00 of the {\tt Cloudy} software, a photoionization code last described by \citet{CLOUDY}. Specifically, we assumed a slab with density $n = 10^{4}$ cm$^{-3}$ and  temperature of $\sim$ 20000~K. The particular choice of these two parameters has been selected after many trials, in order to obtain an accretion spectrum as similar as possible to that of CG98. For the far-UV part of the spectrum ($\lambda \lesssim 3000$~\AA) we follow the results of \citet{France}, and assume a linear decrease of flux at wavelengths shortward of the Balmer continuum. 

\begin{figure}
\centering
\plotone{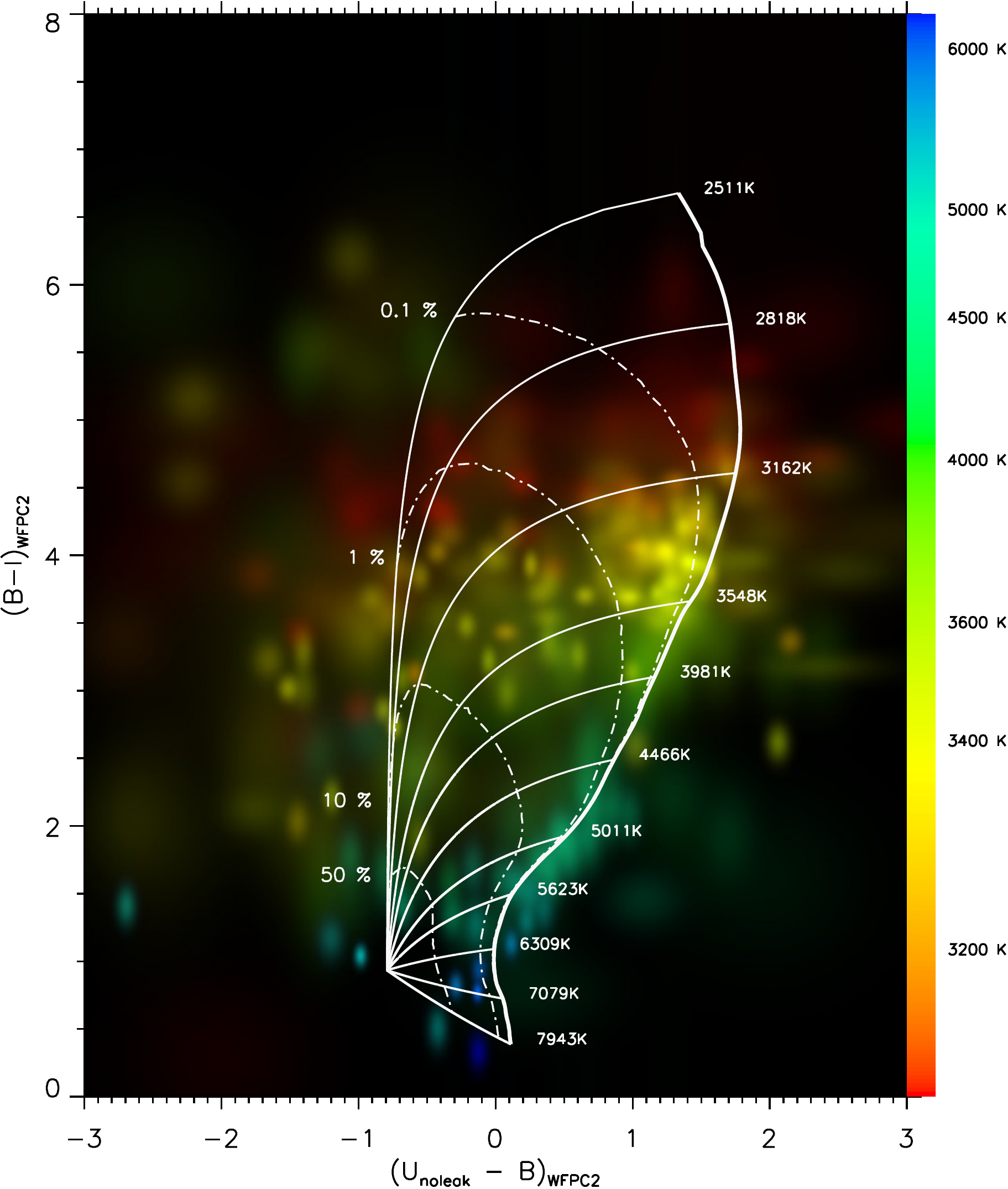}
\caption{2CD of our ONC population. Each source has been corrected for extinction and represented as a normalized 2D gaussian, corresponding to the photometric errors. The sources are color-coded according to their $T_{\rm eff}$, as shown in the scale on the right of the plot.
The thick line represents our calibrated isochrone, for no accretion. The thin lines represent the simulated displacements (for different $T_{\rm eff}$) from the photospheric colors obtained by adding an increasing amount of $L_{\rm acc}/L_{\rm tot}$, assuming the accretion spectrum analogous to that of CG98. Dashed lines indicate constant ratios $L_{\rm acc}/L_{\rm tot}$}
\label{UBI_colored}
\end{figure}

We use our photometrical colors and produce a reddening corrected 2CD (Fig. \ref{UBI_colored}), assuming again the A$_V$ values of \citet{DaRio12}. In this way, the displacements of the ONC members from the isochrone are solely due to accretion and we can investigate this aspect, neglecting the extinction contribution. In this plot sources are represented as equally normalized 2D gaussian, with widths corresponding to the individual photometric error. This facilitates our visual inspection, because, in this representation, sources with small errors appear sharp and bright while sources with poorer photometry are broader and less luminous, contributing to the plot mostly through their relative number. The sources are also color-coded according to their T$_{\rm eff}$, highlighting that the dereddened sources are generally located at bluer colors than the photospheric ones, along stripes that depend on T$_{\rm eff}$.

In Fig.~\ref{UBI_colored} we also show the predicted displacement from the photospheric isochrone, obtained by adding a component of the basic accretion spectrum to the theoretical (calibrated) spectra. Specifically, increasing $L_{\rm acc}/L_{\rm tot}$ from 0 to 1, the colors are moved along lines that start from the theoretical isochrone, at any stellar $T_{\rm eff}$, and converge to the colors of the pure accretion spectrum, located on lower left end of the plot. For this computation, we have used the basic accretion spectrum, analogous to that of CG98. It is evident that, according to this model, accretion affects both the $(B-I)$ and the $(U-B)$ colors, and this prediction is somewhat compatible with the extinction-corrected photometry of some ONC members. 
On the other hand, there are many sources observed at $(U-B)<-1$, a region of this diagram which is not reachable according to this accretion spectrum model. This suggests that for these stars the Balmer jump must be higher than that of our accretion spectrum.

\begin{figure}
\centering
\plotone{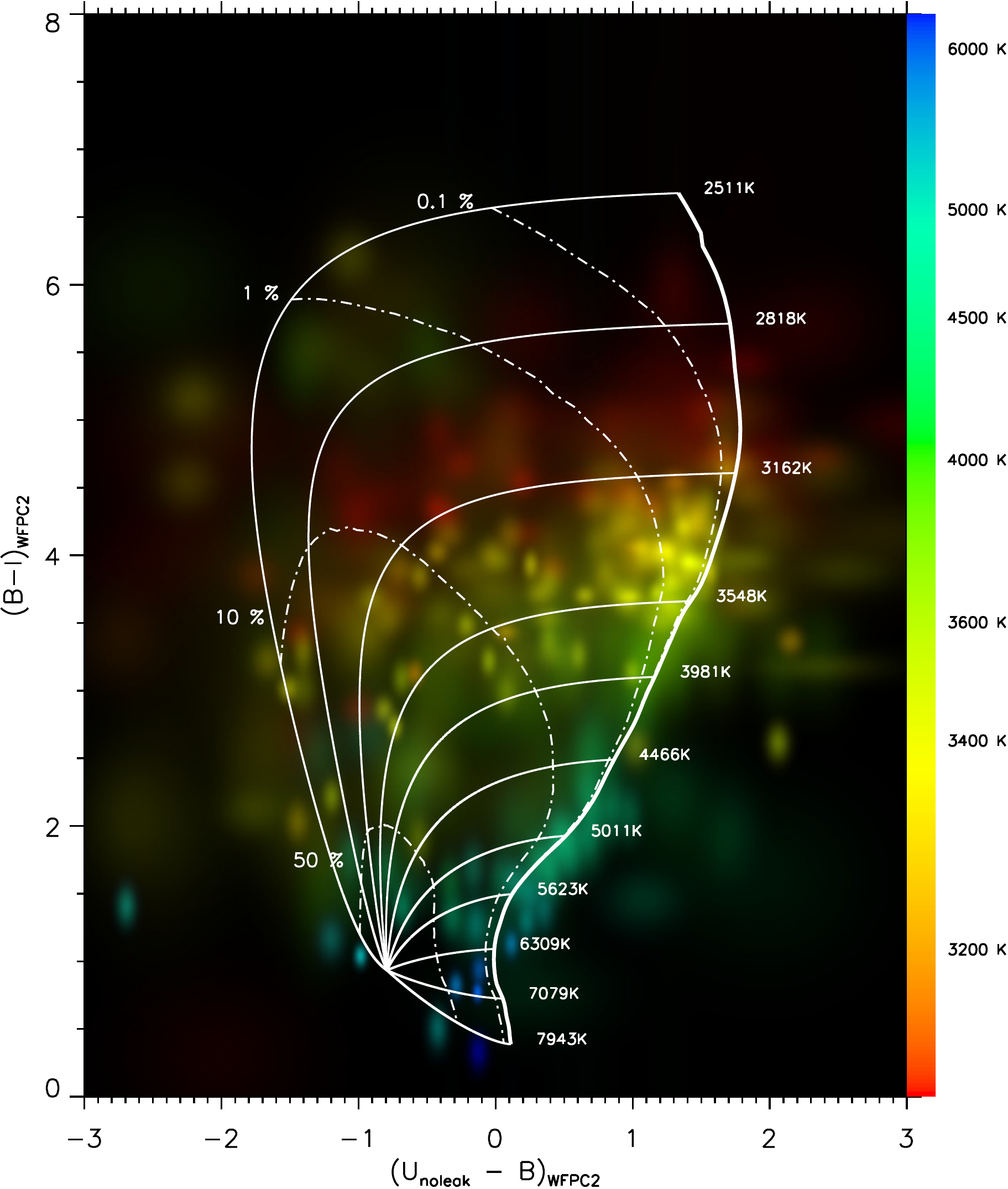}
\caption{Same as Fig.~\ref{UBI_colored}, but adding to the stellar spectra a sum of the accretion spectrum produced used for the model of Fig.~\ref{UBI_colored}, and a pure HII region with decreasing ratio of this component to the first for increasing $L_{\rm acc}$.}
\label{UBI_colored_MIX}
\end{figure}

To correct for the discrepancy we have included another component to this spectrum in order to have a better agreement with our data. This component is modelled with an HII region spectrum, also simulated with {\tt Cloudy}, with the same density as the one previously assumed for the optically thin part but a much lower temperature ($\sim$ 3000 K). This is added to the accretion spectrum in such a way that it becomes relevant only for small amounts of mass accretion. The result is shown in Fig.~\ref{UBI_colored_MIX}: the new model predicts, consistently with our data, a larger displacement in $(U-B)$ for low values of $L_{\rm acc}/L_{\rm tot}$.
For our subsequent analysis we will therefore consider this approach to model the color displacement due to mass accretion.

%-----------------------------------------
\subsubsection{Final Two-Colors Diagram}
\label{MonteCarlo}
From the modeling shown in Fig.~\ref{UBI_colored_MIX}, we derive $L_{\rm acc}/L_{\rm tot}$ for all our ONC sources. For each source, we consider its known $T_{\rm eff}$ from \citet{DaRio12}, and therefore the unique accretion displacement curve in the 2CD that starts from isochrone point corresponding to the stellar temperature of the source. We then consider the observed photometry of the source in the 2CD, and deredden it along the reddening vector until a interception with the accretion curve is found. This intersection defines, geometrically, both the amount of $A_V$ and $L_{\rm acc}/L_{\rm tot}$ for the star. 

\begin{figure*}
\epsscale{1}
\plottwo{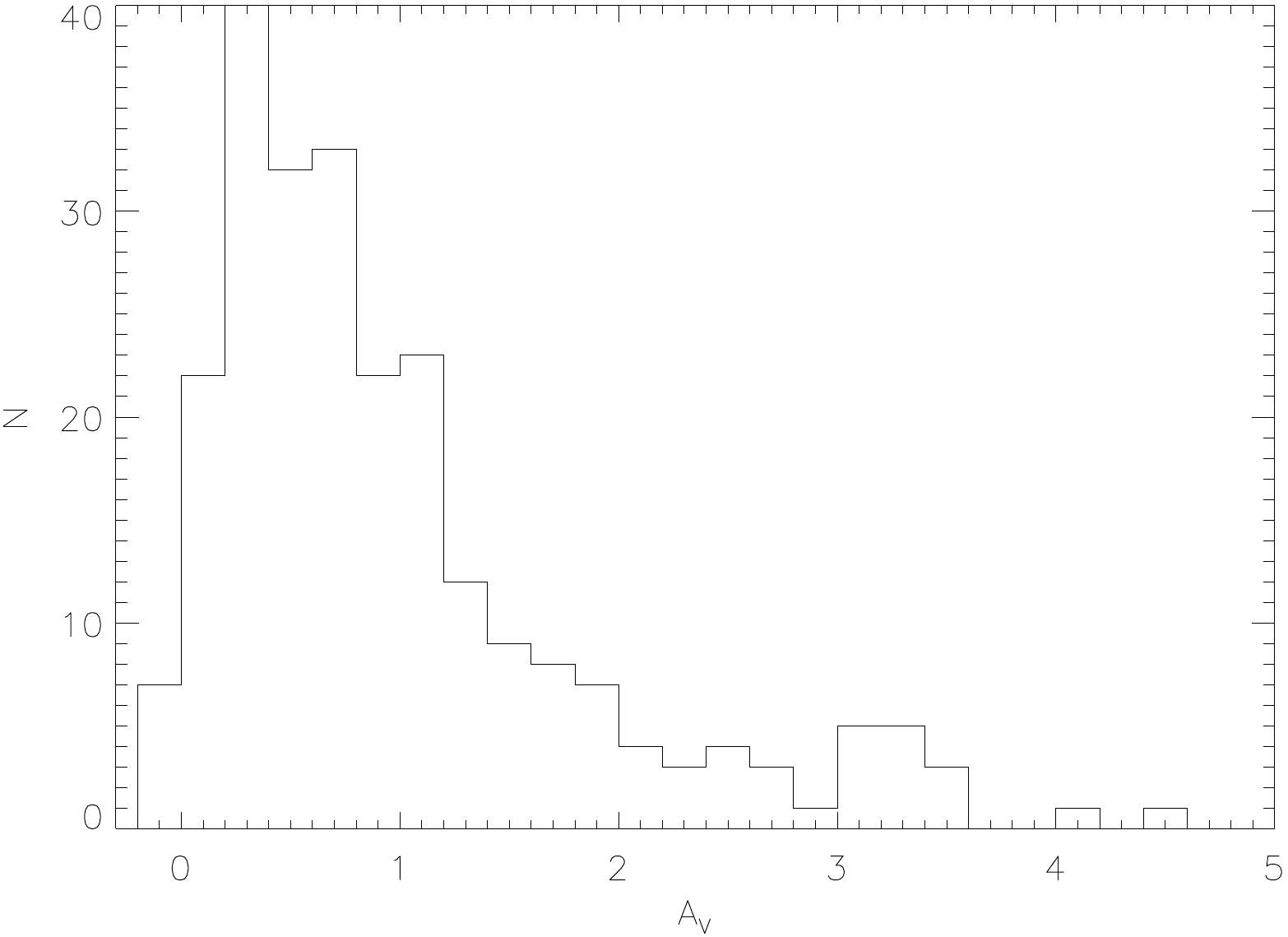}{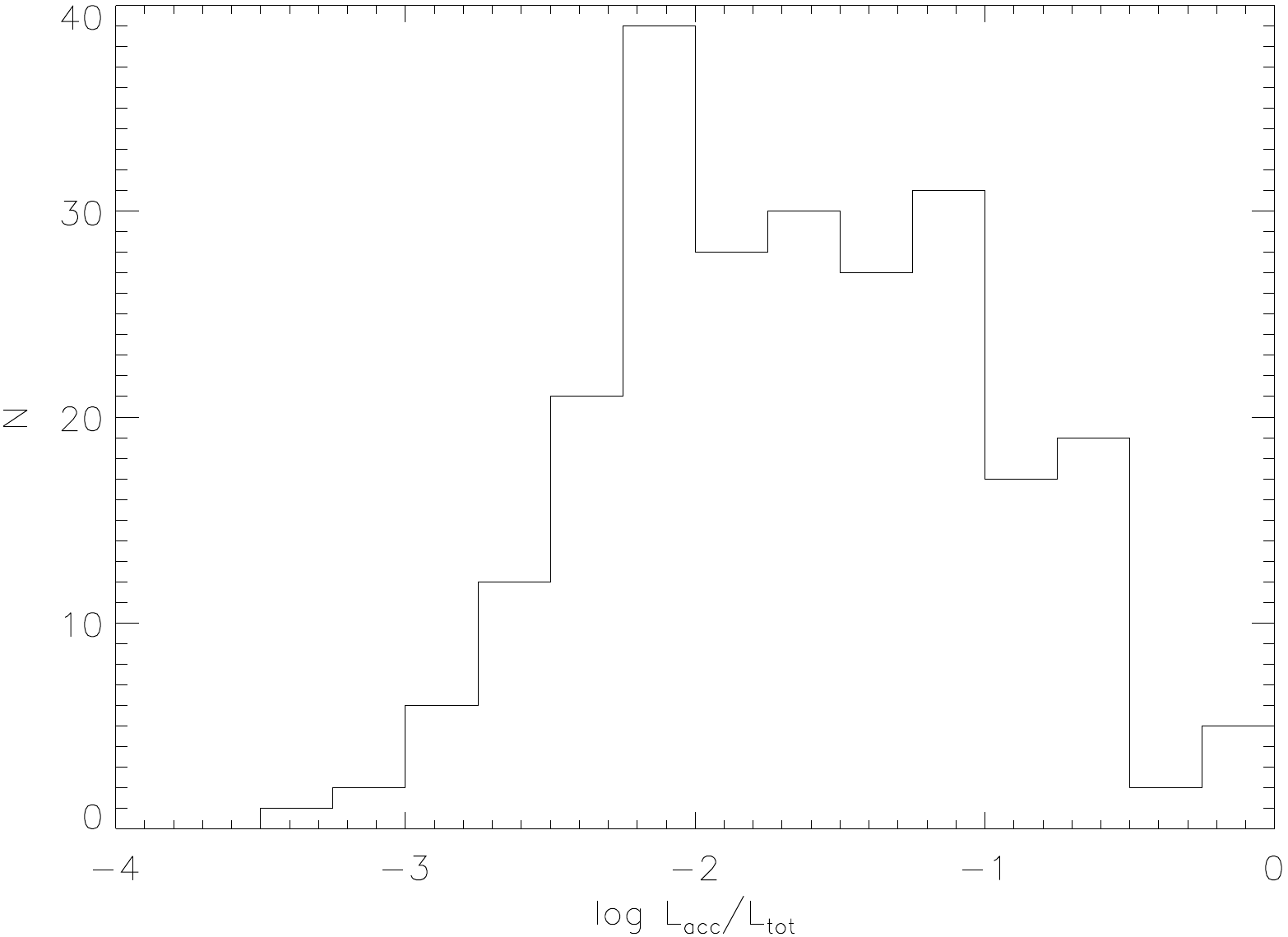}
\caption{Histograms of the values of A$_V$ and L$_{\rm acc}$/L$_{\rm tot}$ estimated for our sources from the analysis of the 2CD. }
\label{geom_meth_res}
\end{figure*}

Using this method, we obtain a solution for 271 sources. In order to quantify the uncertainty in the results, we use a Monte Carlo method: for each source we displace randomly the photometry in each band within the photometric errors, assumed gaussian. We also randomly change the stellar $T_{\rm eff}$ accounting for the uncertainty in the spectral types from spectroscopy, equal to $\pm$1 spectral subtype \citep{H97}. We iterate this approach 1000 times for each star, deriving each time $A_V$ and $L_{\rm acc}/L_{\rm tot}$ values.
From the 1000 results of $A_V$ and $L_{\rm acc}/L_{\rm tot}$ for each source, we determine the 1, 2 and 3 $\sigma$ intervals of the distribution of results. If an intersection is found for all the 1000 iterations, we assume our result to have a 3 sigma confidence; on the other hand if for some of the iterations an intersection is not found (because, e.g., the colors lie outside the color range reachable by the accretion model plus extinction), a lower confidence level is associated to the result. We report these values in Table \ref{2CD_results}. Neglecting sources with confidence lower than 1 $\sigma$, our sample comprises  245 sources. In Fig.~\ref{geom_meth_res} we show the distribution of A$_V$ and L$_{\rm acc}$/L$_{\rm tot}$ for this sample. Our extinctions vary from $\sim$0 to $\sim$ 5, with a mean value of $\sim$ 0.98 $\pm$ 0.06 magnitudes and only 7 sources show a negative value of $A_V$, the lowest being $A_V = -0.18 \pm$ 0.03 mag. We assign to these sources $A_V$ = 0 and retain them in our analysis. Comparing our results with those obtained by \citet{DaRio12} we obtain a very good match, measuring of a distribution of differences $\Delta A_V = A_{V_{\rm this\ work}} - A_{V_{\rm DaRio}}$ peaked at -0.03 mag with a sigma of 0.31 mag. This testifies the accuracy of our method. 

From now on we will use these values of $A_V$ and $L_{\rm acc}$ for all the sources for which this method gave a result. 
On the other hand, a small difference in the estimate of $L_{\rm acc}$ with \citet{DaRio} is expected, given the higher sensitivity of our 2CD to this quantity with respect to their analysis. 

The Monte Carlo approach allows to estimate the errors on our parameters. We derive individual errors from the standard deviation of the values obtained; these will be used in our subsequent analysis to estimate the errors of the \macc . 

\subsection{H$\alpha$ Luminosity}
Besides the $U-$band excess, magnetospheric infall accretion processes lead to line emission, in particular the hydrogen recombination lines \citep[see e.g.,][]{Muzerolle01}. Specifically, the H$\alpha$ excess is often used as an indicator of mass accretion, and its intensity can be empirically related to $L_{\rm acc}$.

As anticipated in Sec.~\ref{selection_sources}, our WFPC2 catalog includes 1342 sources with measured $H\alpha$ photometry. In order to correctly estimate the Equivalent Width of the $H\alpha$ emission (EW$_{H\alpha}$) one requires an estimate of the photospheric continuum at its wavelength. This generally depends depends on a nearby broad band flux (e.g., the $I$-band luminosity), as well as both $T_{\rm eff}$ and $A_V$ of the sources. To this purpose, we consider temperatures from \citet{DaRio12} and extinctions from either our 2CD analysis (Sec.~\ref{MonteCarlo}), or from \citet{DaRio12} for the sources with no results from the 2CD. These required parameters reduce our sample to a total of 1026 sources. The photospheric continuum is then estimated by means for synthetic photometry, computing the photospheric (H$\alpha-I$) color for each star given its $T_{\rm eff}$, and subtracting it from the dereddened $I$ magnitudes. The H$\alpha$ excess, measured from the difference between the observed H$\alpha$ flux and the expected photospheric flux, is then converted in terms on equivalent width, or in terms of fraction of total stellar luminosities. We also stress that the WFPC2 F656N filter is broad enough (FWHM$\sim$28.5 \AA $\sim$ 1300 km/s) to contain the entire H$\alpha$ excess, since this line has a typical broadening of $\sim$200-700~km~s$^{-1}$ ($\sim$4.5-13\AA) \citep[e.g.,][]{WB03,Mortier11}, much smaller than the filter width.

From all the derived values, we consider in our analysis only sources with 3 $<|$EW$_{H\alpha}|<$ 1000. This choice allows us to exclude non-accreting sources, or members showing a suspiciously high excess, probably due to uncertainties or contamination from circumstellar material. We also exclude from our analysis sources that are known from visual inspection on the HST images to be proplyds, binaries or edge-on disks, as already explained in Sec.~\ref{selection_sources}. Ultimately, the final sample of sources with available H$\alpha$ luminosity includes 682 ONC members. We report these results in Table \ref{Ha_results}.

%-----------------------------------------
\subsection{Stellar Luminosity}
\label{stellar_luminosity}

\begin{deluxetable*}{ccccccccc}
\tablecaption{Stellar parameters of the sources where the 2CD method leads to a determination of $A_V$ and L$_{\rm acc}$}
\tablehead{\colhead{OM$^{\rm a}$} & \colhead{RA} & \colhead{DEC} & \colhead{$T_{\rm eff}$$^{\rm b}$}  & \colhead{A$_V$} & \colhead{log(L$_*$/L$_\odot$)} & \colhead{R$_*$} & \colhead{$\Delta I_{\rm acc}$} & \colhead{Confidence$^{\rm c}$} \\
\colhead{ } & \colhead{(J2000)} & \colhead{(J2000)} & \colhead{(K)}  & \colhead{(mag)} & \colhead{  } & \colhead{(R$_\odot$)} & \colhead{(mag)} & \colhead{$\sigma$}}
\startdata
20 & 5:34:15.087 & -5:22:59.971 &         3328 & 0.56 & -1.62 & 0.47 & -0.0020 &            2 \\
30 & 5:34:17.299 & -5:22:47.977 &         3104 & 1.09 & -1.51 & 0.61 & -0.0044 &            1 \\
47 & 5:34:20.794 & -5:23:29.131 &         3388 & 0.51 & -1.01 & 0.90 & -0.0023 &            3 \\
50 & 5:34:24.769 & -5:22:10.443 &         3465 & 0.22 & -0.82 & 1.09 & -0.0021 &            3 \\
63 & 5:34:26.501 & -5:23:23.986 &         3138 & 0.95 & -0.80 & 1.35 & -0.0001 &            1 \\
69 & 5:34:28.962 & -5:23:48.085 &         3107 & 0.42 & -1.61 & 0.54 & -0.0031 &            3 \\
70 & 5:34:29.446 & -5:23:37.437 &         3138 & 0.08 & -1.58 & 0.55 & -0.0002 &            1 \\
77 & 5:34:32.658 & -5:21:7.4565 &         3632 & 0.91 & -0.57 & 1.31 & -0.0197 &            3 \\
87 & 5:34:30.477 & -5:21:55.949 &         3124 & 1.09 & -1.63 & 0.52 & -0.0056 &            1 \\
146 & 5:34:48.830 & -5:23:17.984 &         3705 & 1.29 & -0.16 & 2.03 & -0.0004 &            3 \\
167 & 5:34:50.496 & -5:23:35.331 &         3435 & 0.08 & -0.83 & 1.09 & -0.0364 &            3 \\
... & ... & ... & ... & ... & ... & ... & ...
\enddata
\tablecomments{$^{\rm a}$Orion Master Catalog entry number \citep{HST_treas}. $^{\rm b}$From \citet{DaRio12} and references therein. $^{\rm c}$Sigma of confidence introduced in Sec. \ref{MonteCarlo}. (This table is available in its entirety in a machine-readable form in the online journal. A portion is shown here for guidance regarding its form and content.)}
\label{2CD_results}
\end{deluxetable*}

In order to convert the obtained values of L$_{\rm acc}$/L$_{\rm tot}$, an estimate of the stellar luminosity of our sources is required.
We derive the bolometric luminosity ($L_*$) of each source from the observed $I$ magnitudes, corrected for the effects of extinction and accretion derived in Sec.~\ref{MonteCarlo}.
In particular, accretion is taken into account by computing the excess in magnitudes $\Delta I_{\rm acc}$  in the $I-$band due to the derived amount of accretion for each source. These values $\Delta I_{\rm acc}$ (reported in Table \ref{2CD_results}) have a mean of $\sim -0.04$~mag, a rather small value, since $I$ band is weakly affected by accretion (e.g. \citealt{Fischer}). Thus, $L_*$ is then derived from:
\begin{eqnarray}
\log\left(\frac{L_*}{L_{\odot}}\right) = 0.4 \cdot \left[\textbf{M}_{\rm bol,\odot} - \textbf{M}_{\rm bol,*}\right] \nonumber \\
= 0.4 \cdot [\textbf{M}_{I_{\rm WFPC2},\odot} - \textbf{M}_{I_{\rm WFPC2},*}\nonumber \\
+ BC_{I_{\rm WFPC2},\odot} - BC_{I_{\rm WFPC2},*}]\\
=0.4 \cdot [\textbf{M}_{V,\odot} - (V - I_{\rm WFPC2})_{\odot} - I_{\rm WFPC2} + \nonumber \\
+ \Delta I_{\rm acc} + A_{I_{\rm WFPC2}}+BC_{I_{\rm WFPC2}}(T_{\rm eff})+DM], \nonumber
\end{eqnarray}
where $I_{WFPC2}$ refers to magnitudes in the WFPC2 photometric system, $V$ indicates the $V$-band magnitude in the standard Johnson photometry. We assume the value $M_{V,\odot}$ = 4.83 \citep{Binney}, checking that using as a reference for the solar spectrum the synthetic spectrum of \citet{BT-Settle} with $T_{\rm eff, \odot}$=5780 K and $\log g_\odot$=4.43 we obtain the same result, and we compute through synthetic photometry $(V - I_{WFPC2})_{\odot} = 0.70$. We assume a distance for the ONC of $d$ = 414 $\pm$ 7 pc \citep{distanzaONC}, corresponding to a distance modulus $DM = 8.085$. Extinction is converted from $A_V$ to $A_{I_{WFPC2}}$ using the reddening law of \citet{Cardelli}; also here the values of $A_V$ adopted are those obtained with the 2CD, if available, and from \citet{DaRio12}, when there are no results from the 2CD. Finally, $BC_{I_{\rm WFPC2}}(T_{\rm eff})$ are the I-band bolometric corrections, derived in Sec.~\ref{model_calibration}. The derived $L_*$ are reported in Table \ref{2CD_results} in the cases where $A_V$ is estimated with the 2CD and in Table~\ref{Ha_results} in the others.

Errors in $L_*$ are also derived with a Monte Carlo procedure, similarly to Sec.~\ref{MonteCarlo}, propagating the estimated uncertainties of the photometry, extinction, DM and $\Delta I_{\rm acc}$.

\begin{deluxetable*}{ccccccccc}
\tablecaption{Stellar parameters of the sources where we assumed $A_V$ from \citet{DaRio12} and we estimate L$_{\rm acc}$ from the $H\alpha$ luminosity}
\tablehead{\colhead{OM$^{\rm a}$} & \colhead{RA} & \colhead{DEC} & \colhead{$T_{\rm eff}$$^{\rm b}$}  & \colhead{A$_V$$^{\rm b}$} & \colhead{log(L$_*$/L$_\odot$)} & \colhead{R$_*$} & \colhead{EW$_{H\alpha}$} & \colhead{log(L$_{H\alpha}$/L$_\odot$)}\\
\colhead{ } & \colhead{(J2000)} & \colhead{(J2000)} & \colhead{(K)}  & \colhead{(mag)} & \colhead{  } & \colhead{(R$_\odot$)} & \colhead{(\AA )} & \colhead{  }}
\startdata
  80 & 5:34:33.569 & -5:22:8.7996 &         3165 & 1.32 & -0.80 & 1.33 &   110.40 & -3.40 \\
         101 & 5:34:40.859 & -5:22:42.345 &         4060 & 1.15 &  0.25 & 2.71 &    80.72 & -2.03 \\
         104 & 5:34:41.952 & -5:21:32.102 &         3079 & 1.24 & -1.13 & 0.96 &    31.52 & -4.33 \\
         108 & 5:34:41.614 & -5:23:57.485 &         3106 & 0.80 & -0.86 & 1.29 &     6.83 & -4.71 \\
         111 & 5:34:42.477 & -5:22:46.243 &         3059 & 1.54 & -1.13 & 0.97 &    70.33 & -4.00 \\
         117 & 5:34:41.816 & -5:21:49.500 &         2930 & 0.49 & -1.23 & 0.95 &     8.79 & -5.12 \\
         130 & 5:34:44.797 & -5:22:38.543 &         2954 & 0.83 & -1.70 & 0.54 &    53.61 & -4.79 \\
         133 & 5:34:46.783 & -5:21:29.147 &         3495 & 0.88 & -0.88 & 0.99 &    26.31 & -3.92 \\
         134 & 5:34:45.953 & -5:22:50.690 &         3052 & 2.58 & -1.11 & 0.99 &    11.37 & -4.78 \\
         135 & 5:34:45.987 & -5:22:47.540 &         3034 & 2.60 & -1.33 & 0.79 &   232.10 & -3.71 \\
         144 & 5:34:46.559 & -5:23:25.585 &         3560 & 3.40 & -0.39 & 1.67 &    92.06 & -2.85 \\
          ... & ... & ... & ... & ... & ... & ... & ... & ...
\enddata
\tablecomments{$^{\rm a}$Orion Master Catalog entry number \citep{HST_treas}. $^{\rm b}$From \citet{DaRio12} and references therein.(This table is available in its entirety in a machine-readable form in the online journal. A portion is shown here for guidance regarding its form and content.)}
\label{Ha_results}
\end{deluxetable*}

%_____________________________________
\section{Results: Stellar Properties}
\label{results}
%_____________________________________

\subsection{Accretion Luminosity}

\subsubsection{L$_{\rm acc}$ from the Two-Colors Diagram}
We convert the derived $L_{\rm acc}/L_{\rm tot}$ from the 2CD (Sec.~\ref{MonteCarlo}) in terms of $L_{\rm acc}/L_{\odot}$. This is achieved considering the stellar luminosities $L_*$ derived in Sec.~\ref{stellar_luminosity}, from the relation:
\begin{eqnarray}
\frac{L_{\rm acc}}{L_{\odot}} = \frac{L_{\rm acc}/L_{\rm tot}}{1-(L_{\rm acc}/L_{\rm tot})} \cdot \frac{L_*}{L_{\odot}}.
\end{eqnarray}
In Table \ref{2CD_results_Lacc} we report these values, for a total sample of 245 sources.

\subsubsection{L$_{\rm acc}$ from L$_{H\alpha}$}
\begin{figure}
\centering
\epsscale{1}
\plotone{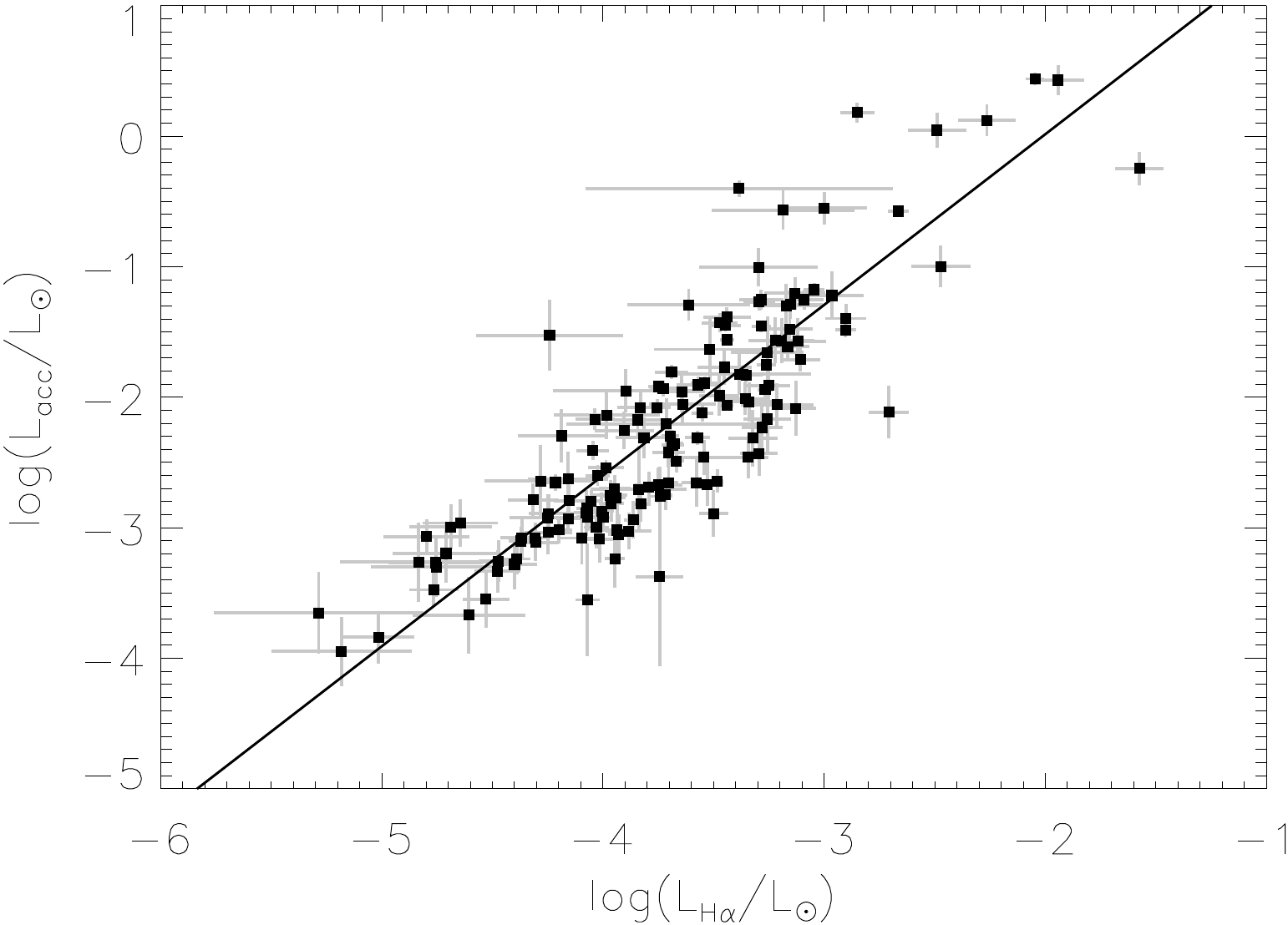}
\caption{Calibration of the $H\alpha$ luminosity as an accretion indicator using the accretion luminosity (L$_{\rm acc}$) obtained with the 2CD. The linear regression fit shown in figure has an analytic expression reported in the text in Eq. \ref{Ha_calib_eq} that permits to obtain accretion luminosity from the $H\alpha$ luminosity estimate.}
\label{Halpha_calibration}
\end{figure}

There are several empirical relations between the $L_{H\alpha}$ and the total accretion luminosity. Since, however, we have an independent estimate of the $L_{\rm acc}$ for some of our ONC sources from the 2CD, we rederive this transformation based on our data. To this purpose, we isolate the sources where both $L_{\rm acc}$ obtained with the 2CD and $L_{H\alpha}$ are available. In particular, limiting to sources with at least a 2 $\sigma$ confidence (see Sec.~\ref{MonteCarlo}), we isolate a sample of 148 stars. Fig.~\ref{Halpha_calibration} shows the relation between these two quantities; to account for the overall uncertainties, including the intrinsic scatter in the values in the $H\alpha$ flux due to variability \citep[e.g.][]{Murphy11}, we add quadratically to all the estimated errors on $\log(L_{\rm acc}/L_\odot)$ a minimum value of 0.2, and use these combined errors for the linear regression. Our fit provides the following relation:
\begin{equation}
\log \left( \frac{L_{acc}}{L_\odot} \right) = (1.31\pm0.03) \cdot \log \left( \frac{L_{H\alpha}}{L_\odot} \right) + (2.63\pm 0.13).
\label{Ha_calib_eq}
\end{equation}

\begin{deluxetable*}{ccccccccccc}
\tablecaption{Accretion values of the sources where the 2CD method leads to a determination of $A_V$ and L$_{\rm acc}$,  according to different evolutionary models}
\tablehead{\colhead{OM$^{\rm a}$} & \colhead{log(L$_{\rm acc}$/L$_\odot$)} & \colhead{M$_*$$^{\rm b}$} & \colhead{log Age$^{\rm b}$}  & \colhead{log \macc $^{\rm b}$}  & \colhead{M$_*$$^{\rm c}$} & \colhead{log Age$^{\rm c}$}  & \colhead{log \macc $^{\rm c}$}  & \colhead{M$_*$$^{\rm d}$} & \colhead{log Age$^{\rm d}$}  & \colhead{log \macc $^{\rm d}$}\\
\colhead{ } & \colhead{  } & \colhead{(M$_\odot$)} & \colhead{(yr)}  & \colhead{(M$_\odot$/yr)} & \colhead{(M$_\odot$)} & \colhead{(yr)}  & \colhead{(M$_\odot$/yr)} & \colhead{(M$_\odot$)} & \colhead{(yr)}  & \colhead{(M$_\odot$/yr)}}
\startdata
          20 & -3.26 & 0.25 & 7.4 & -10.40 & 0.20 & 7.4 & -10.30 & 0.23 & 7.4 & -10.36 \\
          30 & -2.95 & 0.17 & 6.9 &  -9.79 & 0.12 & 6.8 &  -9.66 & 0.13 & 6.7  & -9.69 \\
          47 & -2.60 & 0.25 & 6.5 &  -9.45 & 0.28 & 6.8 &  -9.49 & 0.25 & 6.5  & -9.45 \\
          50 & -2.42 & 0.29 & 6.4 &  -9.25 & 0.33 & 6.6 &  -9.31 & 0.31 & 6.4  & -9.28 \\
          63 & -3.59 & 0.16 & 6.1 & -10.07 & 0.19 & 6.4 & -10.15 & 0.14 & 5.9 & -10.00 \\
          69 & -3.17 & 0.17 & 7.0 & -10.06 & 0.13 & 6.9 &  -9.94 & 0.14 & 6.9  & -9.98 \\
          70 & -4.13 & 0.18 & 7.0 & -11.04 & 0.14 & 7.0 & -10.94 & 0.14 & 6.8 & -10.92 \\
          77 & -1.45 & 0.40 & 6.3 &  -8.33 & 0.42 & 6.5 &  -8.36 & 0.43 & 6.4  & -8.36 \\
          87 & -2.99 & 0.17 & 7.1 &  -9.91 & 0.13 & 7.0 &  -9.77 & 0.14 & 6.9  & -9.83 \\
         146 & -2.31 & 0.32 & 5.7 &  -8.92 & 0.47 & 6.1 &  -9.08 & 0.47 & 6.0  & -9.09 \\
         167 & -1.56 & 0.27 & 6.4 &  -8.37 & 0.31 & 6.6 &  -8.42 & 0.28 & 6.4  & -8.38 \\
 ... & ... & ... & ... & ... & ... & ... & ... & ... & ... & ...
\enddata
\tablecomments{$^{\rm a}$Orion Master Catalog entry number \citep{HST_treas}. $^{\rm b}$Obtained with \citet{D'Antona} models. $^{\rm c}$Obtained with \citet{Siess} models. $^{\rm d}$Obtained with \citet{Palla} models.(This table is available in its entirety in a machine-readable form in the online journal. A portion is shown here for guidance regarding its form and content.)}
\label{2CD_results_Lacc}
\end{deluxetable*}

\citet{Herczeg08}, fitting the Balmer Jump and veiling features and correlating those estimate for $L_{\rm acc}$ to the spectroscopical estimated EW$_{H\alpha}$, found a relation with similar slope $(1.20\pm0.16)$. \citet{Dahm} measured a slope of $(1.18\pm0.26)$ using 14 members of the Taurus-Auriga complex, combing previous $L_{\rm acc}$ estimate with their spectroscopical EW$_{H\alpha}$. \citet{DeMarchi} assumed a unitary slope, implying the proportionality between $L_{H\alpha}$ and $L_{\rm acc}$, and derived another relation using the same data of \citeauthor{Dahm}. We rely on our estimate, and we thus use the relation in Equation \ref{Ha_calib_eq} to obtain $L_{\rm acc}$ for 528 additional sources with no estimate of this quantity from the 2CD.

\begin{deluxetable*}{ccccccccccc}
\tablecaption{Accretion values of the sources where we assumed $A_V$ from \citet{DaRio12} and we estimate L$_{\rm acc}$ from the $H\alpha$ luminosity, according to different evolutionary models}
\tablehead{\colhead{OM$^{\rm a}$} & \colhead{log(L$_{\rm acc}$/L$_\odot$)} & \colhead{M$_*$$^{\rm b}$} & \colhead{log Age$^{\rm b}$}  & \colhead{log \macc $^{\rm b}$}  & \colhead{M$_*$$^{\rm c}$} & \colhead{log Age$^{\rm c}$}  & \colhead{log \macc $^{\rm c}$}  & \colhead{M$_*$$^{\rm d}$} & \colhead{log Age$^{\rm d}$}  & \colhead{log \macc $^{\rm d}$}\\
\colhead{ } & \colhead{  } & \colhead{(M$_\odot$)} & \colhead{(yr)}  & \colhead{(M$_\odot$/yr)} & \colhead{(M$_\odot$)} & \colhead{(yr)}  & \colhead{(M$_\odot$/yr)} & \colhead{(M$_\odot$)} & \colhead{(yr)}  & \colhead{(M$_\odot$/yr)}}
\startdata
          80 & -1.82 & 0.17 & 6.1 &  -8.33 & 0.21 & 6.4 &  -8.42 & 0.16 & 6.0 &  -8.29 \\
         101 & -0.03 & 0.40 & 5.4 &  -6.60 & 0.75 & 5.9 &  -6.87 & 0.80 & 5.8  & -6.90 \\
         104 & -3.04 & 0.16 & 6.3 &  -9.66 & 0.15 & 6.6 &  -9.63 & 0.12 & 6.2  & -9.54 \\
         108 & -3.54 & 0.15 & 6.1 & -10.02 & 0.18 & 6.4 & -10.09 & 0.13 & 6.0  & -9.94 \\
         111 & -2.61 & 0.15 & 6.3 &  -9.21 & 0.14 & 6.6 &  -9.18 & 0.11 & 6.2  & -9.07 \\
         117 & -4.08 & 0.12 & 6.4 & -10.60 & 0.10 & 6.6 & -10.51  & $< 0.10$ & ...   &  ... \\
         130 & -3.65 & 0.11 & 6.8 & -10.34 &  $< 0.10$ & ... &    ...  & $< 0.10$ & ...   &  ... \\
         133 & -2.50 & 0.31 & 6.5 &  -9.40 & 0.34 & 6.7 &  -9.44 & 0.33 & 6.6  & -9.43 \\
         134 & -3.64 & 0.15 & 6.3 & -10.21 & 0.14 & 6.6 & -10.17 & 0.11 & 6.2 & -10.07 \\
         135 & -2.22 & 0.15 & 6.5 &  -8.91 & 0.11 & 6.7 &  -8.76 & 0.10 & 6.3  & -8.74 \\
         144 & -1.11 & 0.31 & 5.9 &  -7.78 & 0.38 & 6.2 &  -7.87 & 0.37 & 6.2  & -7.86 \\
 ... & ... & ... & ... & ... & ... & ... & ... & ... & ... & ...
\enddata
\tablecomments{$^{\rm a}$Orion Master Catalog entry number \citep{HST_treas}. $^{\rm b}$Obtained with \citet{D'Antona} models. $^{\rm c}$Obtained with \citet{Siess} models. $^{\rm d}$Obtained with \citet{Palla} models.(This table is available in its entirety in a machine-readable form in the online journal. A portion is shown here for guidance regarding its form and content.)}
\label{Ha_results_Lacc}
\end{deluxetable*}

\subsection{Stellar Masses and Ages}
\label{HRD_sec}

To derive the \macc \ for our sources, we still need to estimate the masses of our sources ($M_*$); this is accomplished using evolutionary model interpolation of the position of our sources on the Hertzprung-Russel diagram (HRD). Moreover, this procedure gives us the age of the sources that will be used in the later analysis.

\begin{figure}
\epsscale{1.1}
\plotone{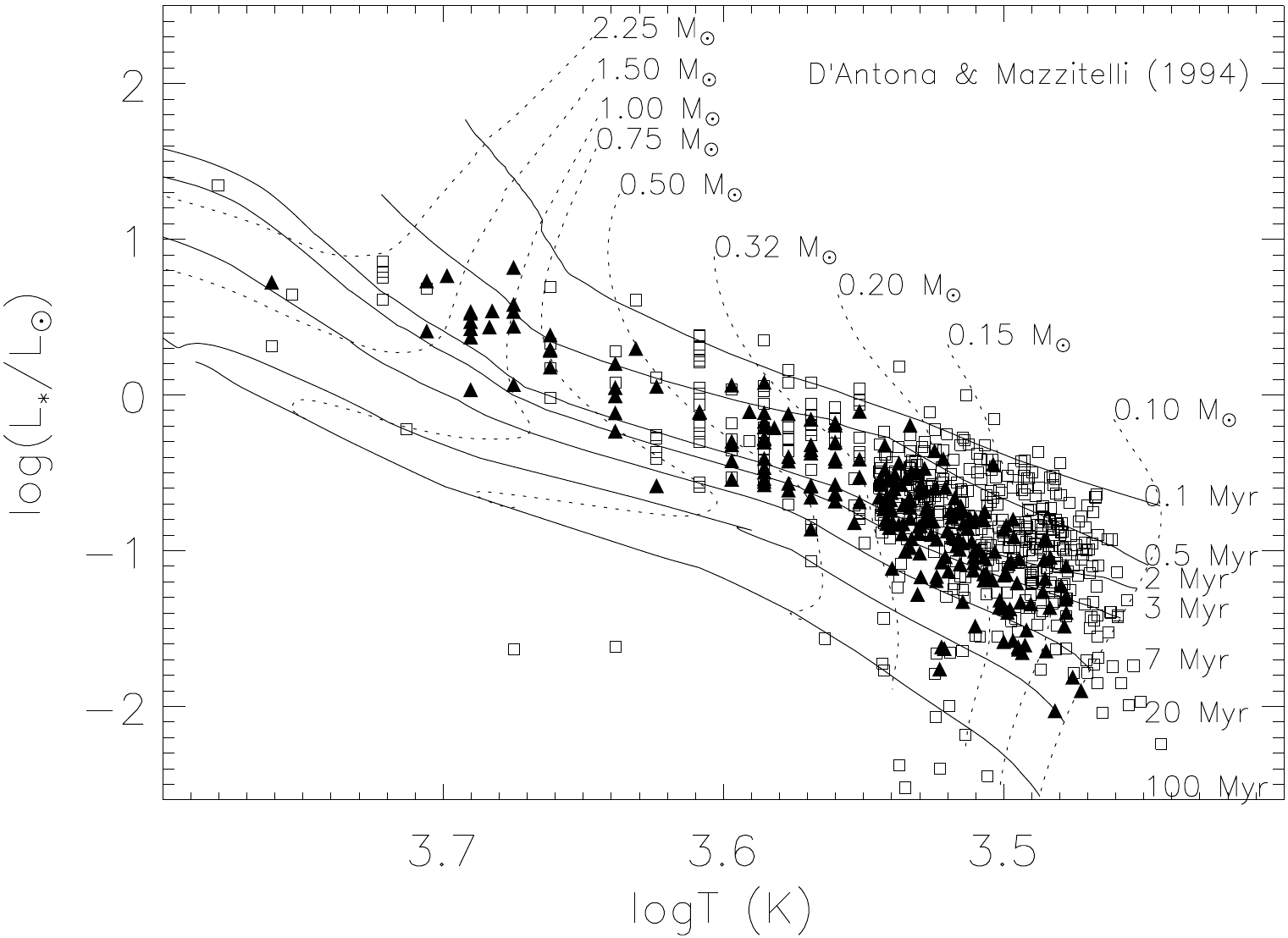}
\plotone{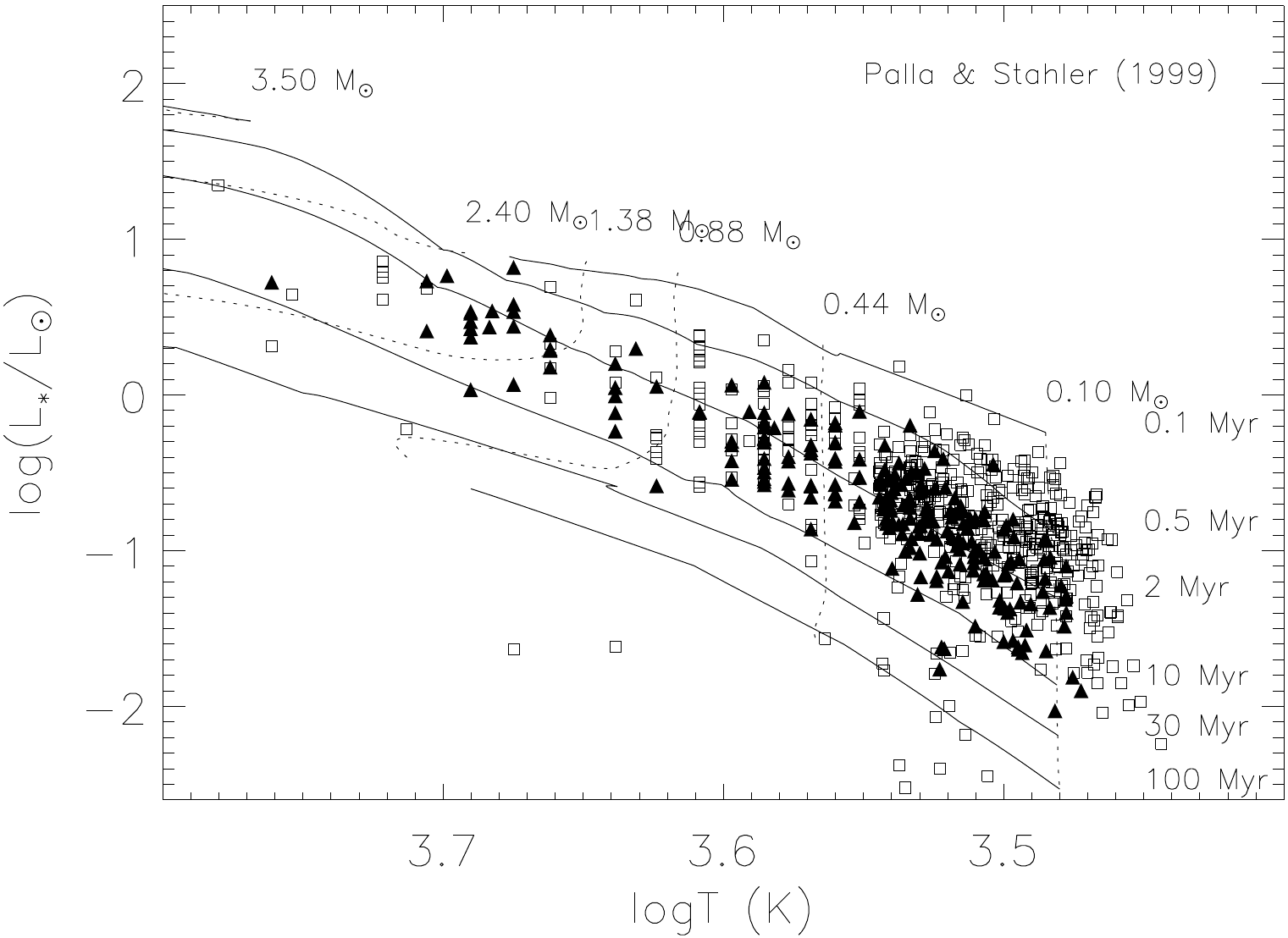}
\plotone{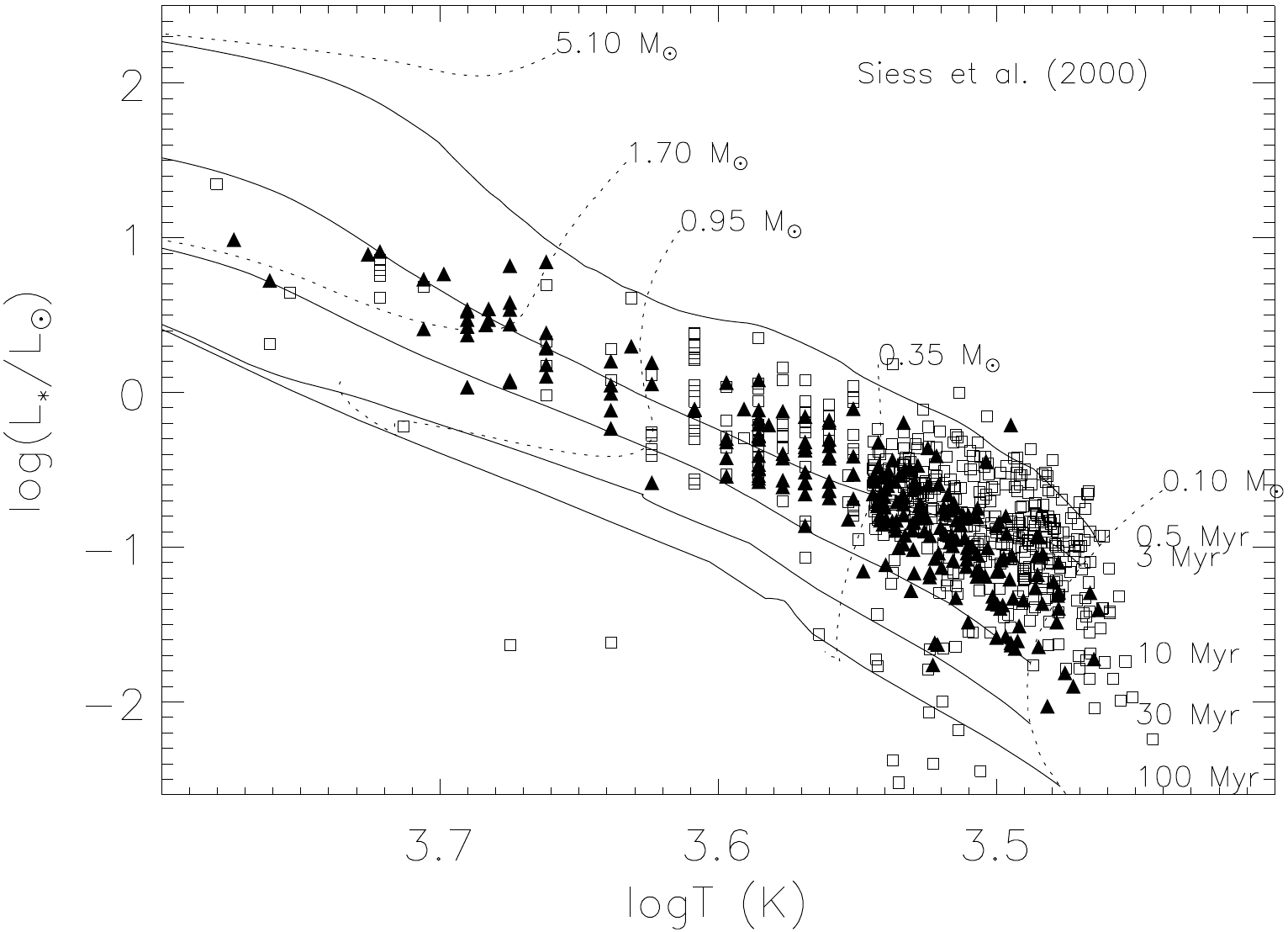}
\caption{The Hertzsprung-Russel Diagram of our ONC sources with overplotted different PMS evolutionary models. Filled triangles refer to sources whose $L_{\rm acc}$ has been estimated using the 2CD, while empty squares are sources with results derived from the $L_{H\alpha}$.}
\label{HRD-evol}
\end{figure}

In Fig.~\ref{HRD-evol} we present the HRD of our sources, built using $T_{\rm eff}$ from \citet{DaRio12} and the luminosity ($L_*$) derived in Sec.~\ref{stellar_luminosity}.
Sources with a measurement of accretion from the 2CD are plotted in triangles, whereas accretion estimates from the H$\alpha$ excess are represented by open squares. From Fig.~\ref{HRD-evol} it is evident that our sample of \macc\ estimates extends well down to the Hydrogen-burning limit. In the very-low mass range, our sample includes members at all luminosities (or ages), in agreement with the overall luminosity spread measured in Orion \citep{DaRio,DaRio12}.
We find some sources located on our HRD below the Main Sequence; these are  either relatively less luminous sources seen in scattered light \citep{Skemer11} or non-members of the ONC, and will be excluded from our analysis.

\begin{figure*}
\epsscale{0.9}
\plottwo{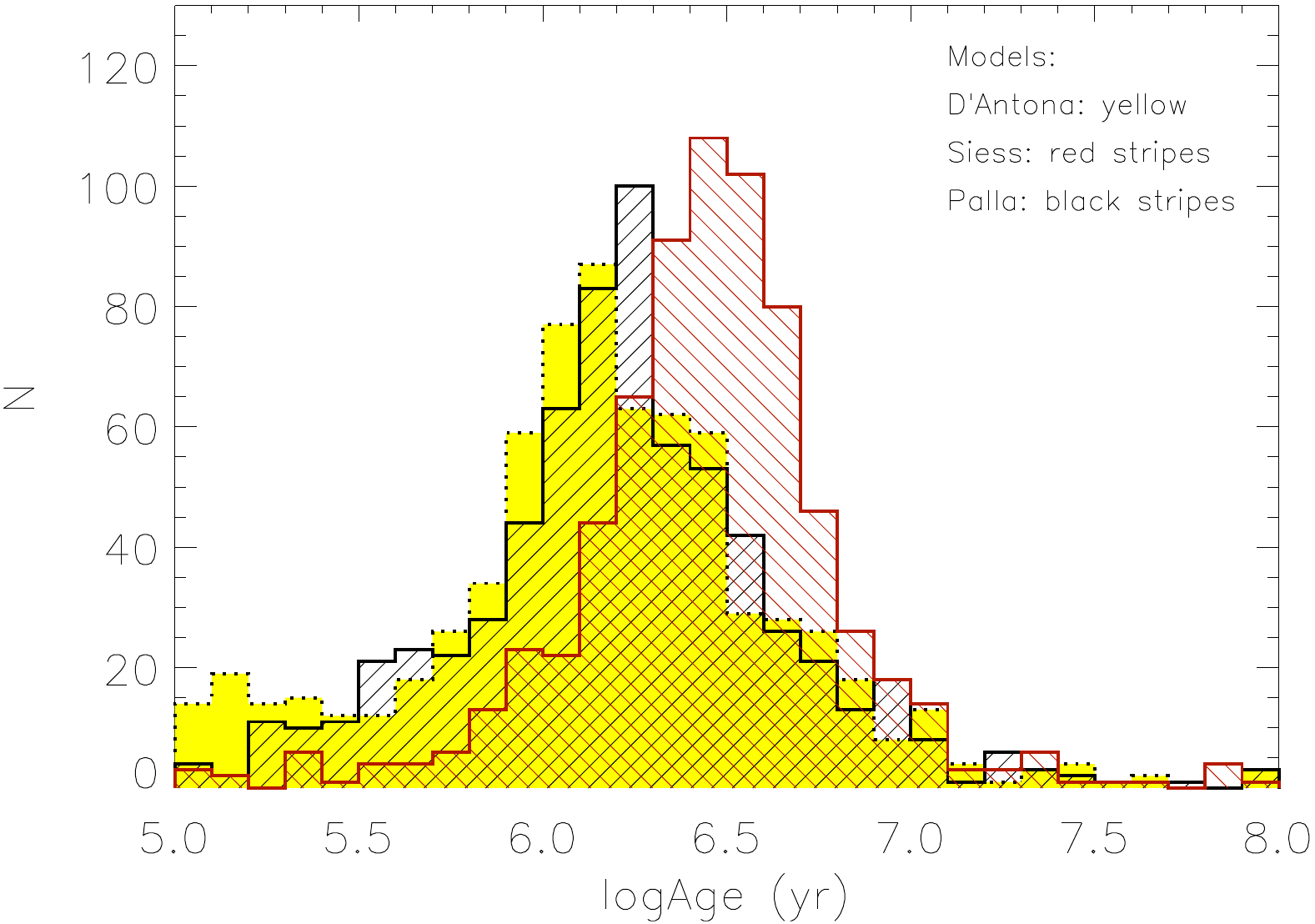}{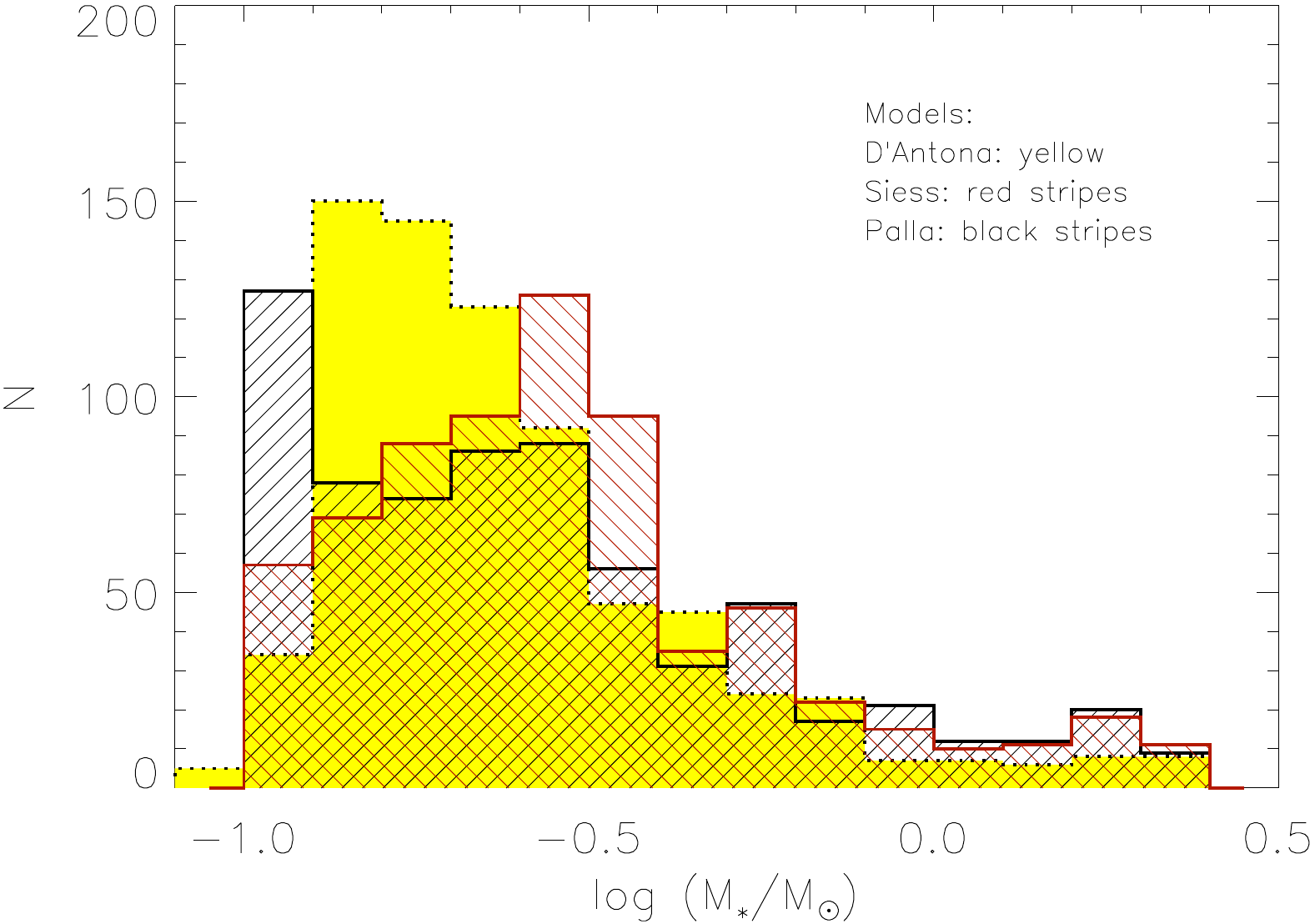}
\caption{The age ({\em left panel}) and mass ({\em right panel}) distributions for our stellar sample, according to the three family of evolutionary models we have considered.}
\label{sample}
\end{figure*}

We assign masses and ages to our sources by interpolating theoretical isochrones and evolutionary tracks on the HRD. We use models of \citet{D'Antona}, \citet{Siess} and \citet{Palla} (see Fig.~\ref{HRD-evol}). In Fig.~\ref{sample} we present the age and mass distributions obtained from the different evolutionary models.

Considering only sources that appear on the HRD at higher $L_*$ than the ZAMS, our stellar sample includes 730 sources with $M_*$ and age estimates from \citeauthor{D'Antona}, 675 using \citeauthor{Palla} models, and 697 assuming isochrones from \citeauthor{Siess}

\subsubsection{Completeness analysis}
\begin{figure}
\epsscale{1}
\plotone{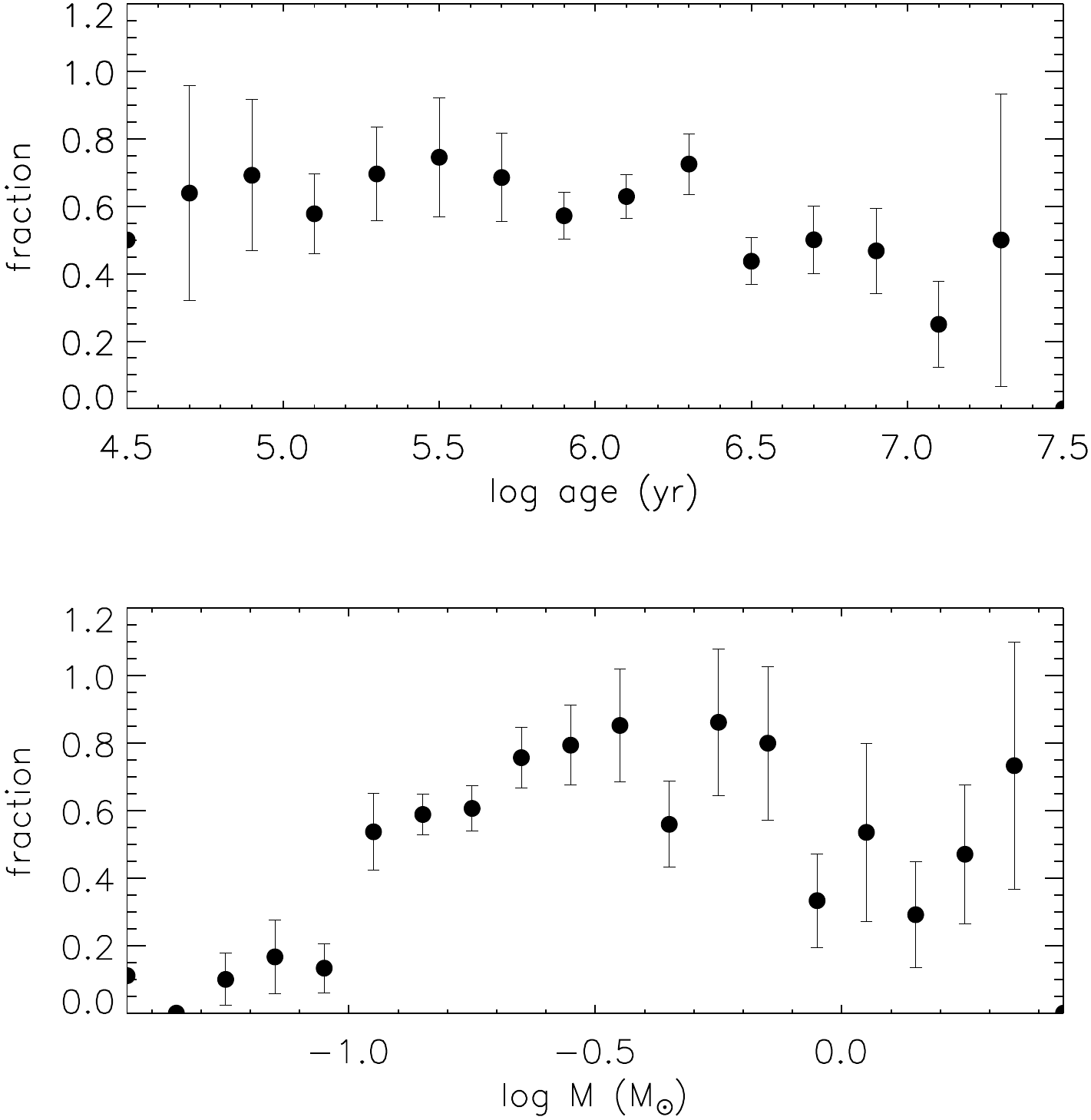}
\caption{Fraction of the sources in our samples with respect to the sample of \citet{DaRio12}, that is complete at 100\% down to the H-burning limit. We see that we have a $\sim$70\% of completeness down to the H-burning limit.}
\label{completeness_fig}
\end{figure}

The completeness of these samples depends on two factors: the availability of $T_{\rm eff}$ from \citet{DaRio12} and the detection of the sources in $U$ or H$\alpha$ (within the WFPC2 field of view). Whereas the $U-$band photometry tends to be shallow for highly reddened stars, the $H\alpha$ photometry is much less affected by extinction, and extends well into the brown dwarf mass range. Therefore, source detection is not expected to alter the representativeness of our stellar sample. On the other hand, the available $T_{\rm eff}$ spectral types from \citet{DaRio12} are characterized by a uniform and close to 100\% completeness down to the H-burning limit. Therefore, we expect our sample of \macc \ to be representative of the ONC stellar population down to $\sim 0.1M_\odot$, and the lower number of sources in our catalog compared to \citet{DaRio12} is mainly due to our smaller field of view.
To test this hypothesis, and rule out significant selection effects in our sample of \macc\ estimates, we consider full catalog of available stellar parameters from \citet{DaRio12}, limited to the same (smaller) field of view of our WFPC2 survey. Then we compute, as a function of mass and age, the fraction of these sources included in our \macc\ sample. The result, obtained assuming \citeauthor{D'Antona} models, is shown in Fig.~\ref{completeness_fig}. The error bars represent the poissonian uncertainty from the stellar numbers in both samples. As supposed, we do not detect significant trends in the fraction of accreting sources with respect to the stellar paramters, except for a significant lack of sources below 0.1~M$_\odot$. Above this mass, about 2/3 of the ONC members have a \macc\ measurement, regardless their mass or age.

\hbox{}

\subsection{Mass Accretion Rates}
As show in Equation~\ref{Macc_eq},  \macc\ is derived from the measured total accretion luminosity $L_{\rm acc}$, the stellar radius $R_*$ and the mass $M_*$. We obtain the radius of the sources from their $T_{\rm eff}$ and $L_*$. 
The obtained values of \macc \ are reported in Table \ref{2CD_results_Lacc} and Table \ref{Ha_results_Lacc}, separately for source with $L_{\rm acc}$ estimated from the 2CD and from the H$\alpha$ excess.

\begin{figure}
\epsscale{0.9}
\plotone{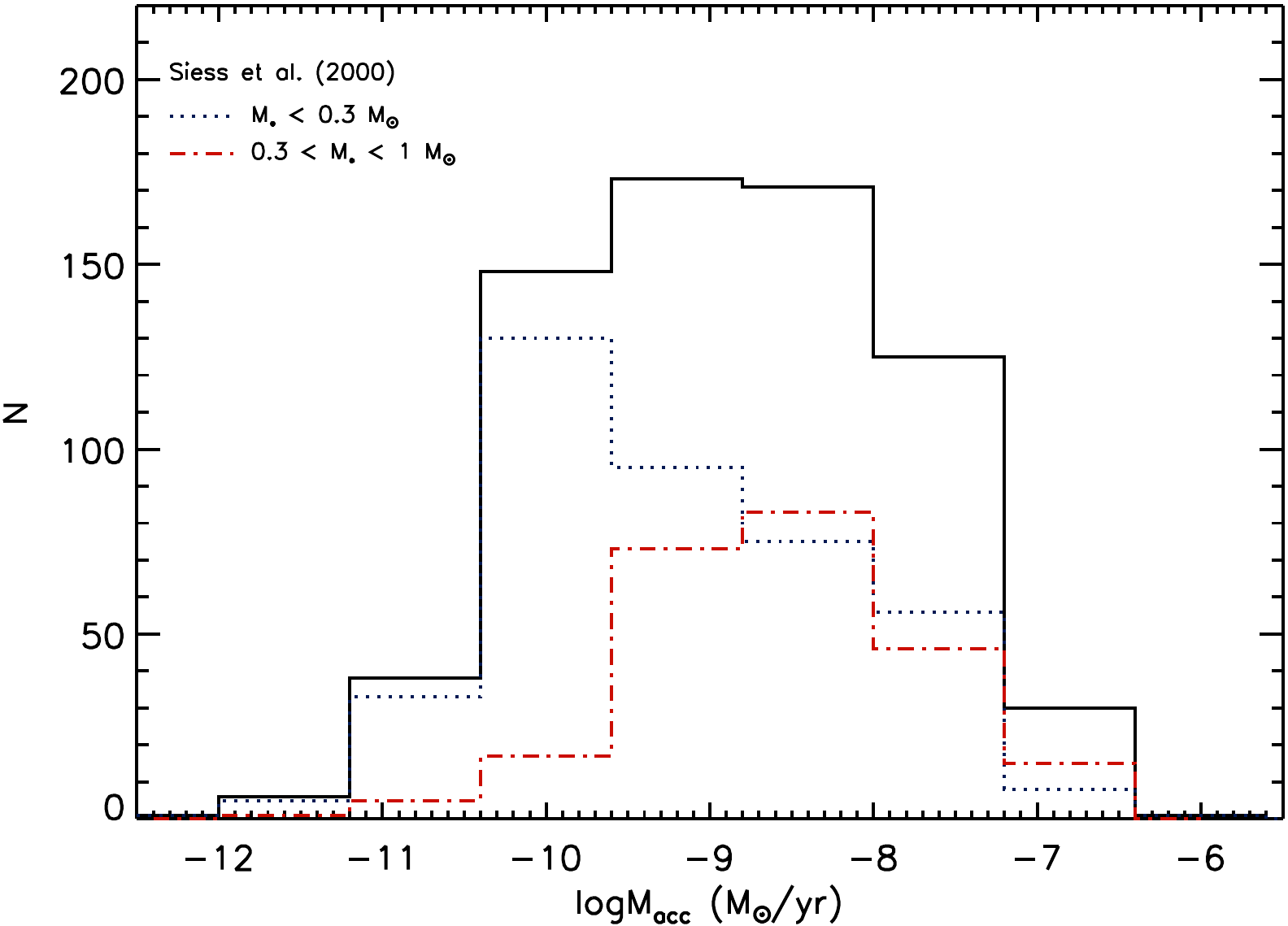}\\
\plotone{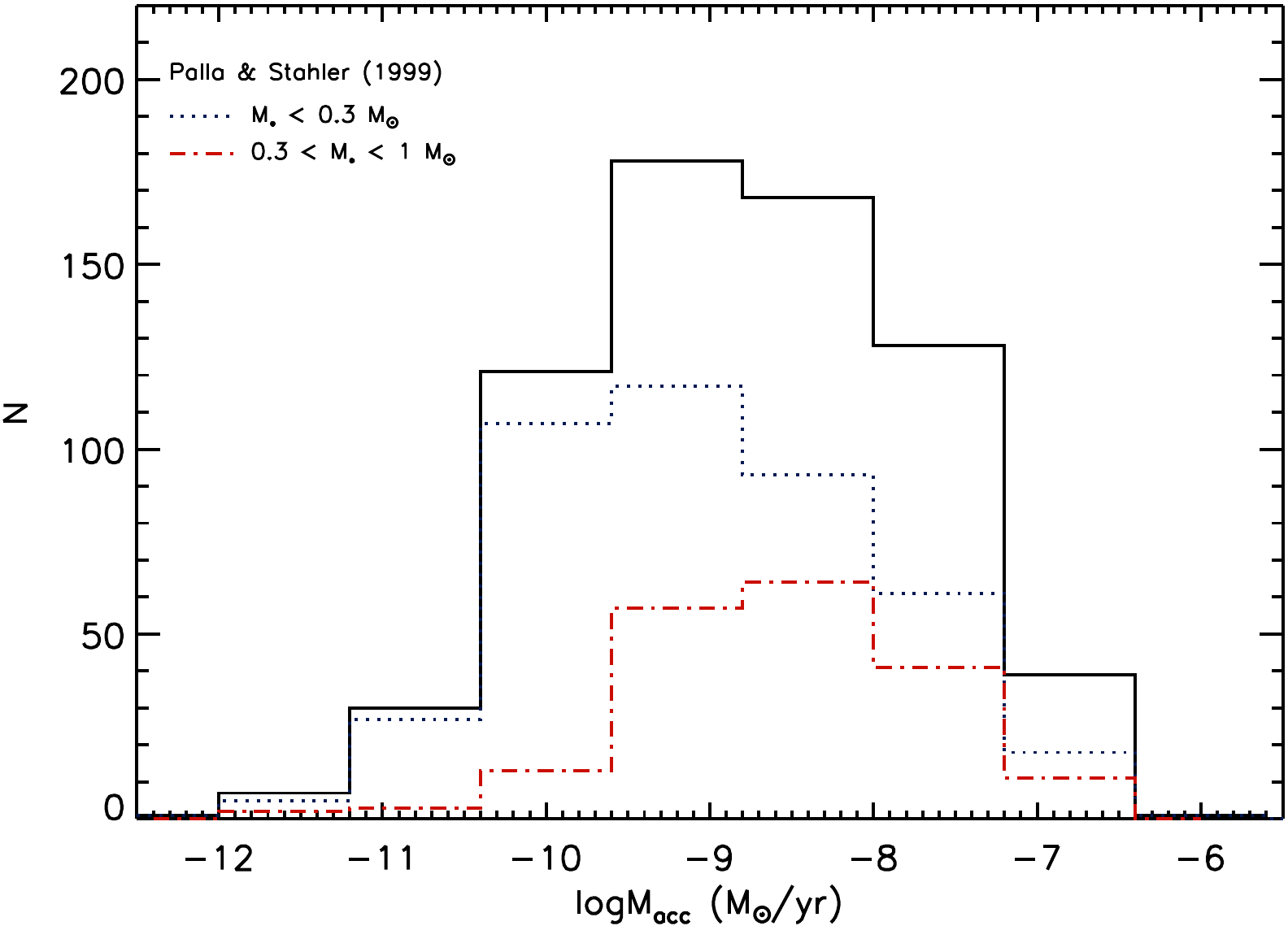}\\
\plotone{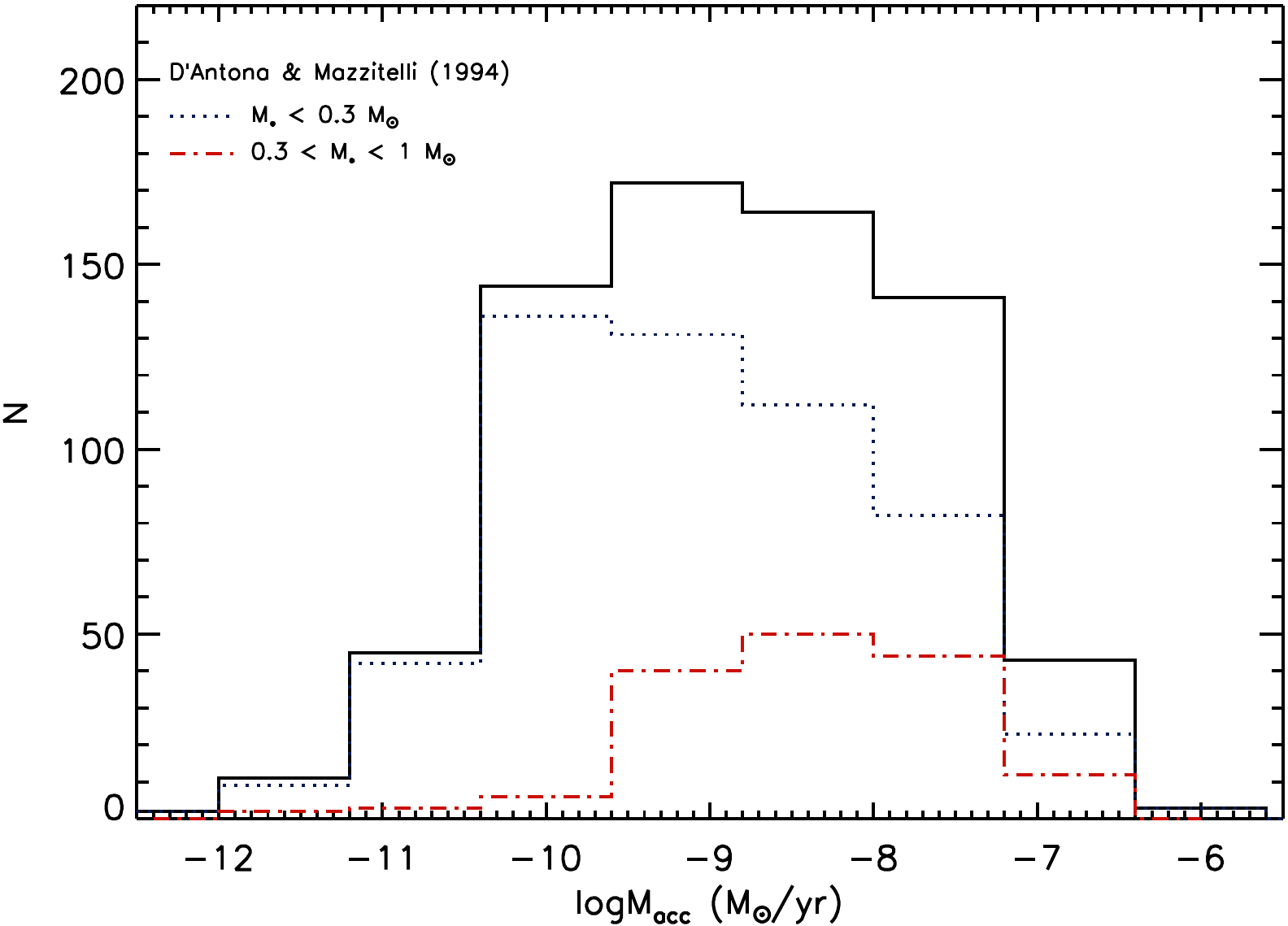}
\caption{Distribution of mass accretion rates according to the three different evolutionary models. In each plot we also present the partial distributions separating the sample in two mass bins, highlighting the different averages.}
\label{Macc_histo}
\end{figure}
Fig.~\ref{Macc_histo} shows the distribution of \macc\ according to the three evolutionary models. The distributions appear similar to each other, spanning a range between $\sim 10^{-11}$ and $\sim 10^{-7}$ $M_{\odot}$/yr, with a mean value of about $\sim 1.4 \cdot 10^{-9} M_{\odot}$/yr. This is compatible with what was previously found for the Trapezium region \citep{Robberto}.

In the same plot we also compare the distributions separately for very low mass stars ($M_*<0.3$M$_{\odot}$, blue dotted histograms) and low- and intermediate mass stars ($M_*>0.3$M$_{\odot}$, red dot-dashed histograms).
It is evident that, regardless of the evolutionary model, intermediate mass stars show higher mass accretion rates than lower mass stars. In the next section we will investigate this trend in more detail.

\subsection{L$_{\rm acc}$ from Censored Data}
\label{censored_data}
Our WFPC2 catalog of ONC sources provides photometric upper limits \citep{HST_treas} for sources undetected in one or more photometric bands; this allows us to obtain, for some sources, additional upper values of $L_{\rm acc}$.
For 55 sources with available $T_{\rm eff}$ and detection in $B$ and $I$, but undetected in $U$-band, we could derive upper limits on $L_{\rm acc}$ with the 2CD method. For 8 of these stars with no correspondent $L_{H\alpha}$ estimate, we find an intersection in the 2CD, leading to an upper limit on their $L_{\rm acc}$ that can be used in our analysis. These values are also reported in Table \ref{2CD_results_Lacc}. For other 21 of these sources with intersection on the 2CD, $L_{\rm acc}$ was already derived from the $L_{H\alpha}$ excess, therefore, as a sanity check, we compare these values with the upper limits obtained from $U$: in 14 cases (66\%) we find that the upper limit value is higher than the derived one, while in the remaining 7 (33\%) cases the two results are compatible.

For 84 sources we have $H\alpha$-band upper limit values and corresponding spectral types, therefore we are able to derive the upper limit of $L_{\rm acc}$ from Equation \ref{Ha_calib_eq}. These values are reported in Table \ref{Ha_results_Lacc}.

%_______________________________________________________
\section{Analysis: Dependence of L$_{\rm acc}$ and \macc\ on Stellar Parameters}
\label{analysis}

\subsection{$L_{\rm acc}$ vs $L_*$}
\begin{figure}
\centering
\epsscale{1.1}
\plotone{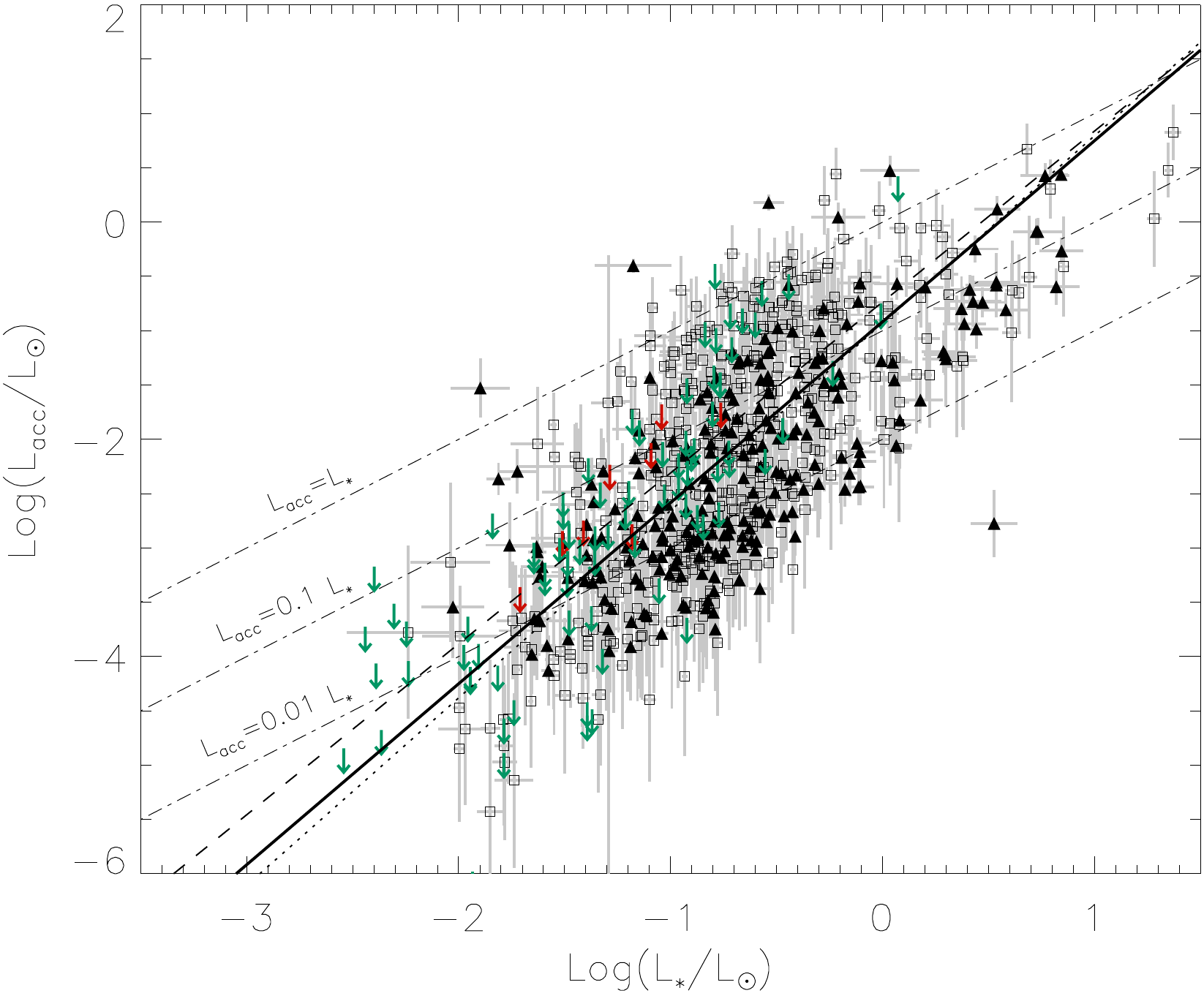}
\caption{Accretion luminosity vs stellar luminosity. Values obtained from the 2CD are plotted as triangles, while squares represent data obtained from $H\alpha$ excess. Dot - dashed lines represent different ratio between accretion and stellar luminosity and the solid line is our best fit (slope$\sim$1.68) from a linear least-squares regression which accounts for errors on both L$_*$ and L$_{\rm acc}$. The fit assuming only results from the 2CD is showed with a dotted line, while the one from only $H\alpha$ luminosities with a dashed line.}
\label{LaccvsL_geom_Ha}
\end{figure}

In Fig.~\ref{LaccvsL_geom_Ha} we plot $L_{\rm acc}/L_{\odot}$ as a function of $L_*/L_\odot$. We stress that these two quantities are independent of the assumed evolutionary model, allowing a less biased analysis of the dependence of accretion with the stellar properties.
Because of this, such relation has been long investigated: \citet{Clarke06}, reporting results from \citet{Natta06}, showed an almost uniform distribution of sources on this plot between the upper bound $L_{\rm acc}\sim L_*$ and the lower observational detection threshold, with a relation $L_{\rm acc}\propto L_*^\beta$ with $\beta=1.6$. \citet{Tilling} were able to theoretically reproduce the observed dispersion for $M_* \ge 0.4\ M_\odot$, even if they could not reach the lowest values of $L_{\rm acc}$ detected observationally. Assuming \citeauthor{D'Antona} evolutionary models, they used a dependence of \macc \ $\propto t^{-\eta}$ with $\eta=1.5$ \citep{H98} obtaining $\beta=1.7$. 

Fig.~\ref{LaccvsL_geom_Ha} shows that our plot is widely populated between $L_{\rm acc} \sim L_*$ and $L_{\rm acc} \sim 0.01 \ L_*$, with several sources below this range, down to $L_{\rm acc} \sim 0.001 \ L_*$. 
The upper locus found by \citet{Natta06} is still present at $L_{\rm acc} \sim L_*$ in our sample. We observe 21 sources over this level, but 17 of which are compatible within 2 sigma to $L_{\rm acc} \le L_*$. The remaining 4 sources may be "continuum stars" \citep{Calvet}, i.e. sources where the accretion component dominates over the stellar emission. If this were the case, their spectral types might also be highly uncertain, with also an impossibility to determine the correct $L_*$.

We note also that the density of sources with $L_* \gtrsim L_\odot$ is much lower than that of lower luminosity sources, probably because of observational selection effects. In particular, as will be pointed out in Sec.~\ref{systematics}, for higher mass stars our 2CD does not allow us to detect low values of $L_{\rm acc}$. Viceversa, the lack of high values of $L_{\rm acc}$ for $L_*\lesssim 0.3\ L_\odot$ occurs in a region widely populated by upper limit detections, and thus cannot be attributed to observational limits.

Fitting our results with a linear least-squares procedure, accounting for the errors (corrected as in Sec.~\ref{MonteCarlo} for both stellar and accretion luminosity), we obtain a slope $\beta=1.68$ $\pm$ 0.02. This is also shown in Figure~\ref{LaccvsL_geom_Ha} as a solid line. This value is compatible with those obtained observationally by \citet{Natta06} and theoretically from \citeauthor{Tilling}. We note, however, that the slope slightly changes when selecting only sources with $L_{\rm acc}$ determined from the 2CD, for which $\beta=1.73\pm0.02$ (dotted line), or from $L_{H\alpha}$, for which $\beta=1.59\pm0.04$ (dashed line).

%_______________________________________________________
\subsection{$\dot{M}_{\rm acc}$ vs Age and $M_*$}
\label{MaccvsagevsM_sec}
\subsubsection{Fit to the data}
To investigate the variations of \macc\ with respect to the stellar age and mass, we consider all these 3 quantities in a single a 3D space.
To this purpose, we neglect sources with suspicious ages, specifically, too young ($\log$ (t/yr) $< 5.5$) or too old ($\log$ (t/yr) $> 7.3$). 

\begin{figure}
\epsscale{1}
\plotone{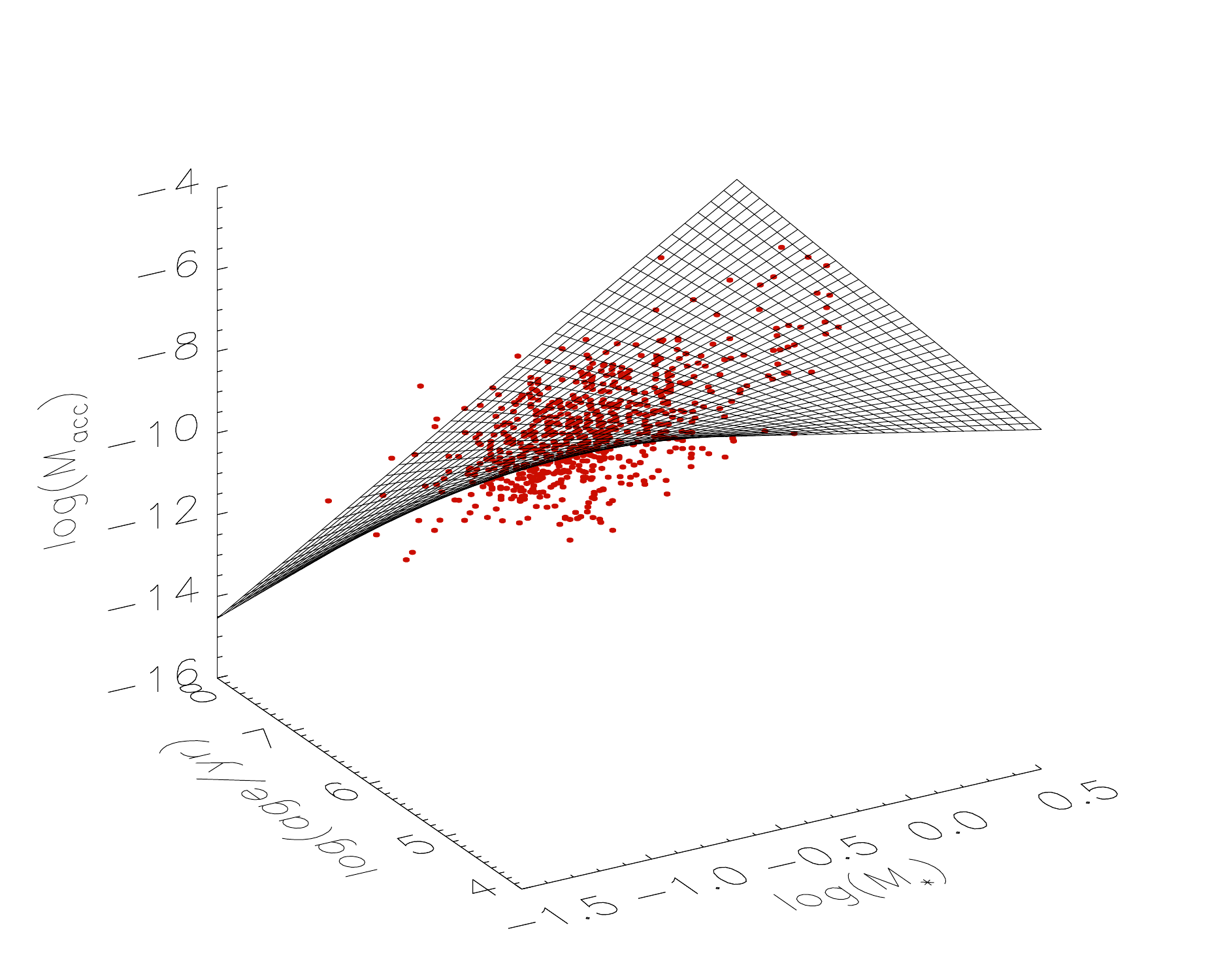}
\caption{Snapshot of our 3D diagram of \macc \ with respect to both $M_*$ and age, together with our best fit surface (see text). The analytical expression of the latter is also reported in Eq.~\ref{fit3D_all} and the coefficients of the fit are reported in Table~\ref{param_fit_table}. The full 3D animation is available electronically. In Fig.~\ref{MaccvsagevsM_panels} different views of this surface are shown.}
\label{fit3D}
\end{figure}

\begin{deluxetable*}{lcccc}
\tabletypesize{\scriptsize}
\tablecaption{Coefficients of Eq. \ref{fit3D_all} obtained fitting the values of \macc \ as a function of both $M_*$ and the age of the sources for different subsamples}
\tablehead{\colhead{Sample}  & \colhead{$A$}  & \colhead{$B$} & \colhead{$C$} & \colhead{$D$}}
\startdata
All sources & $-(5.12 \pm 0.86)$ & $- (0.46 \pm 0.13)$ & $-(5.75 \pm 1.47)$ & $(1.17 \pm 0.23)$ \\
Only 2CD results & $-(4.70 \pm 1.29)$ & $- (0.56 \pm 0.21)$ & $(0.35 \pm 0.35)$ & $-(0.54 \pm 2.20)$ \\
Only $H\alpha$ results & $-(3.29 \pm 1.36)$ & $- (0.71 \pm 0.21) $ & $(0.93 \pm 0.34)$ & $-(4.01 \pm 2.19)$ 
\enddata
\tablecomments{The sample is divided in subsamples according to the method used to obtain L$_{\rm acc}$, and thus \macc .}
\label{param_fit_table}
\end{deluxetable*}

We perform a minimum $\chi ^2_{\rm red}$ fit, accounting for errors on the single values, assuming two different forms: a) a simple plane ($\log$\macc = $A + B\cdot \log t + C\cdot \log M_*$) and b) a warped surface ($\log$\macc = $A + B\cdot \log t + C\cdot \log M_* + D\cdot \log t \cdot \log M_*$). Specifically, the first form is the simplest function to describe correlations of \macc\ with the stellar parameters, the second allows for a mixed term that accounts for a possible different evolution timescale of \macc\ with stellar mass. To perform the fit we use a standard Levenberg-Marquard non-linear regression.
In the two cases the resulting $\chi ^2_{\rm red}$ are comparable ($\sim 7 - 9$), and the second one being slightly smaller. This decrease of $\chi ^2_{\rm red}$ is not surprising, given that the first function is a subcase of the second, the latter having an additional parameter. Therefore, in order to quantitatively decide whether the most complex model is truly more representative of the real distribution of the data, or whether the simplest model is good enough given the uncertainties, we run a statistical F test. This test compares the relative increase in the residual sum of squares obtained by reducing the complexity of the model, with the relative increase of the degrees of freedom. Then, the test derives the probability that the simpler model is suitable enough. In any case, we find this probability lower than 0.1\%, meaning that introducing the fourth parameter D in our model (i.e., a warped surface instead of a flat plane) leads to a more representative match with our data, and this is not a statistical casualness. Our fitted surface is shown in Fig.~\ref{fit3D}, and described by the following equation:

\begin{eqnarray}
\log\dot{M}_{\rm acc}= A + B \cdot \log t + C \cdot \log M_*  + D \cdot \log t \cdot \log M_*,
\label{fit3D_all}
\end{eqnarray}
where the parameters' values are reported in Table \ref{param_fit_table}, also in the cases where we consider only sources with results obtained from the 2CD or with accretion estimated using the $H\alpha$ luminosity.

\begin{figure*}
\centering
\epsscale{0.7}
\plotone{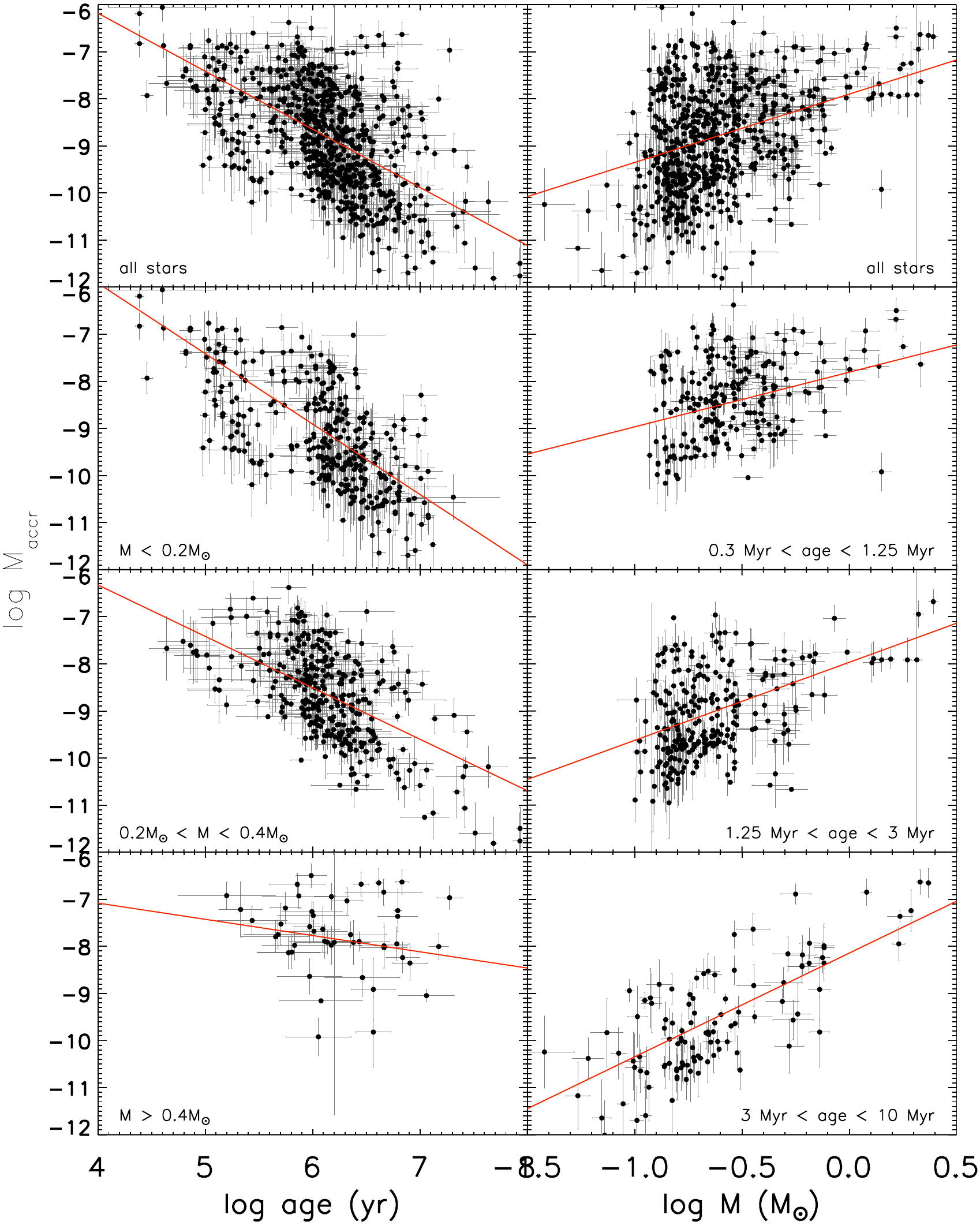}
\caption{\macc \ evolution as a function of individual age and mass of the sources for different ranges of these parameters. The red line is plotted according to the value of the fit obtained using all the sources in the sample (Eq. \ref{fit3D_all}) and for the mean value of the parameter in the range written on the picture. }
\label{MaccvsagevsM_panels}
\end{figure*}

These relations imply that both parameters play together a role in the evolution of mass accretion. To better understand this behavior, and in particular the role of the ``mixed'' term described by the parameter $D$, we separately investigate the dependence of \macc\ on age or mass; also, in every case, we divide the sample in different ranges for these parameters. This is illustrated in Table \ref{Maccvsage_table} for the dependence on age, and in Table \ref{Maccvsmass_table} for the dependence on mass, and the plots are shown in Fig.~\ref{MaccvsagevsM_panels}.

\begin{deluxetable}{lccc}
\tabletypesize{\scriptsize}
\tablecaption{\macc \ vs. age relation coefficients}
\tablehead{\colhead{  }  & \colhead{$\eta$}  & \colhead{$\eta (2CD)$} & \colhead{$\eta (H\alpha)$} }
\startdata
$M_* \sim 0.13 M_\odot$ & $1.50\pm0.26$ & $0.87\pm 0.41$ & $1.53\pm 0.40$ \\
$M_* \sim 0.2 M_\odot$ & $1.28\pm 0.21$ & $0.80\pm 0.32$ & $1.36\pm 0.32$ \\
$M_* \sim 0.3 M_\odot$ & $1.07\pm0.18$ & $0.74\pm 0.28$ & $1.20\pm 0.28$ \\
$M_* \sim 0.5 M_\odot$ & $0.81\pm0.15$ & $0.66\pm 0.23$ & $0.99\pm 0.23$ \\
$M_* \sim 0.8 M_\odot$ & $0.58\pm0.13$ & $0.59\pm 0.21$ & $0.80\pm 0.21$ \\
$M_* \sim 1 M_\odot$ & $0.46\pm0.13$ & $0.56\pm 0.21$ & $0.71\pm 0.21$
\enddata
\tablecomments{Results of the fit \macc \ $\propto t^{-\eta}$ obtained considering all the sources (second column), sources with results obtained with the 2CD (third column) and with the $H\alpha$ luminosity (fourth column). Values of the slope are different for stars belonging to subsamples with different mean $M_*$ (first column).}
\label{Maccvsage_table}
\end{deluxetable}

\begin{deluxetable}{lccc}
\tabletypesize{\scriptsize}
\tablecaption{\macc \ vs. $M_*$ relation coefficients}
\tablehead{\colhead{  }  & \colhead{$b$}  & \colhead{$b (\rm 2CD)$} & \colhead{$b (H\alpha)$} }
\startdata
age $\sim 0.8$ Myr & $1.15\pm 2.00$ & $1.51\pm 3.02$ & $1.50\pm 2.97$ \\
age $\sim 1$ Myr & $1.26\pm 2.02$ & $1.54\pm 3.04$ & $1.59\pm 2.99$ \\
age $\sim 2$ Myr & $1.61\pm2.06$ & $1.65\pm 3.12$ & $1.87\pm 3.06$ \\
age $\sim 5$ Myr & $2.08\pm2.13$ & $1.79\pm 3.22$ & $2.24\pm 3.16$ \\
age $\sim 8$ Myr & $2.32\pm2.16$ & $1.86\pm 3.27$ & $2.43\pm 3.21$ \\
age $\sim 10$ Myr & $2.43\pm2.18$ & $1.89\pm 3.29$ & $2.52\pm 3.23$
\enddata
\tablecomments{Results of the fit \macc \ $\propto M_*^{b}$ obtained considering all the sources (second column), sources with results obtained with the 2CD (third column) and with the $H\alpha$ luminosity (fourth column). Values of the slope are different for stars belonging to subsamples with different mean age (first column).}
\label{Maccvsmass_table}
\end{deluxetable}

%_______________________________________________________
\subsubsection{Discussion}
If we consider only the dependence of \macc\ on stellar ages (\macc $\propto t^{-\eta}$), our fit implies that a similar slope as that proposed by \citet{H98} ($\eta = 1.5$)  is obtained for $M_* \sim 0.13 \ M_\odot$. For higher mass stars we find decreasingly lower values of $\eta$ with mass. Table \ref{Maccvsage_table}, in particular, shows that the values of $\eta$ decrease monotonically regardless of considering all sources, only those with results obtained from the 2CD or with the $H\alpha$ luminosity. Thus, our data indicate that there are different accretion evolutionary timescales for sources with different mass. Surprisingly, for sources with $M_* \gtrsim 0.5 M_\odot$ the slope is smaller than 1. This value is not compatible with the similarity solution model framework \citep{LBP,H98}. Indeed, if we suppose that the disk evolves with time following a simple power law ($M_{\rm disk} \propto t^{-\alpha}$), we have that $\dot{M} = dM_d/dt \propto t^{-(\alpha+1)}$, thus $\eta = \alpha+1 > 1$ in order to have a disk mass decreasing with time.

On the other hand, according to the self-similar models, \macc\ scales with time as a simple power law only in the asymptotic regime, when the age of the system $t\gg t_{\nu}$, where $t_{\nu}$ is the viscous timescale of the disk (set by the initial conditions and by the relevant viscosity). At earlier times, the relationship between \macc \ and $t$ is indeed shallower. Our results might thus be explained within the framework of the self-similar solutions if higher mass stars are not yet in the asymptotic regime, implying that the viscous time is a growing factor of the stellar mass. Note that this trend is the opposite of what suggested by \citet{Alexander06} in the brown dwarf regime to explain the $M_*$-\macc\ correlation.

Alternatively, one might question the validity of the self-similar models themselves in describing, from a statistical point of view, the evolution of protostellar disks, and in particular the assumption that viscosity scales with radius as a simple power law. Indeed, different physical processes are expected to redistribute angular momentum within the disk at different radii, with the magneto-rotational instability (MRI, \citet{MRI}) dominating at small distances from the star, the gravitational instability \citep{Lodato+04} dominating in the colder outer parts, and with the possibility of having extended dead zones \citep{Gammie96} in between the two.

For what concerns the dependence of \macc\ over mass (see Table \ref{Maccvsmass_table}) modeled as \macc $\propto M_*^b$, the trend is similar regardless on the method used to estimate \macc. In particular, the slope in higher, thus the relation is steeper, when considering older ages. This also shows that at older ages more massive sources maintain an higher accretion rate.

Fig.~\ref{MaccvsagevsM_panels} illustrates once more these trends. We plot the evolution of our \macc\ values against age and $M_*$ for different ranges of the parameters. In this figure the slopes of the overplotted lines are obtained from Equation \ref{fit3D_all}, substituting to the projected parameter (mass for the left panels, age for the right ones), the mean value within the subsample used for each particular panel.

The slower temporal decay of \macc \ we measure in the intermediate-mass range ($M_*>$ 0.4 $M_\odot$) is in agreement with estimates of mass accretion rates from $H\alpha$ photometry obtained in the Magellanic Clouds \citep{DeMarchi,DeMarchi11,Spezzi12} and in the massive galactic cluster NGC3603 \citep{Beccari10}. Indeed, in all these studies, the investigated stellar samples do not reach the low-mass regime, due to poor photometric sensitivity. This suggest, as inferred by the authors of these works, that their relatively high values of \macc\ measured for isochronal ages of $\sim$10Myr compared to galactic low-mass stars indicate a selection effect because of higher masses.

%_______________________________________________________
\subsubsection{Caveats}
\label{systematics}
Our derivation of \macc\ using the 2CD may be, in principle, affected by some modest selection effects. This, as mentioned, is located in the fact that for high $T_{\rm eff}$ sources, the color displacement due to accretion is very small, limiting the possibility to identify weak accretors among intermediate mass stars. Specifically, Fig.~\ref{UBI_colored_MIX} shows that for $(B-I)\lesssim 2$, the 1\% $L_{\rm acc}$/$L_{\rm tot}$ line is undistinguishable from the zero accretion one. This is simply because stars with high $T_{\rm eff}$ emit a non-negligible fraction of their flux in the $U$-band, ``hiding'' a possible small accretion excess. As mentioned, however, this problem is highly mitigated since we have also derived $L_{\rm acc}$ from the $L_{H\alpha}$ excess for all the sources for which \macc\ is not derived from the 2CD. In the end, with the H$\alpha$ measurements, and even considering upper limits (see, e.g, Fig.~\ref{LaccvsL_geom_Ha}) the relations we find do not appear to suffer from this mentioned selection effect.

As already introduced in Sec.~\ref{HRD_sec}, the accuracy of the isochronal ages to represent the true stellar ages has been recently questioned. \citet{Baraffe-HRD} and \citet{Baraffe10} claimed that the observed spread of $L_*$ in the HRD of star forming regions can be explained as a consequence of the protostellar phase accretion history. In particular, episodic events of intense accretion (\macc $\sim 10^{-4} M_\odot$/yr) during this phase can produce a large spread of values of stellar radii for stars of identical age and mass. Their results are based on the assumption of "cold" accretion and initial masses $M_{\rm i}$ between 1 $M_J$ and 0.1 $M_\odot$. 
On the other hand, \citet{Hosakawa10} demonstrate that the apparent spread produced by different accretion histories may be important only for sources with $T_{\rm eff} \gtrsim $ 3500 K ($\sim$ 30\% of our sources). In their analysis they assume a fixed $M_{\rm i}$ = 0.01 $M_\odot$, and this could suggest that the results obtained should be revised changing the value of this parameter. Thus, from a theoretical point of view the debate is still ongoing and a unified picture is not yet reached. At the present moment, therefore, we can only use the age inferred from the evolutionary tracks used in the past (e.g. \citet{D'Antona}), waiting for new isochrones based on episodic accretion models.

We stress that if the isochronal ages we assume are inaccurate, because of the aforementioned effects, this results in an overall "horizontal" broadening of our results in the \macc\ vs. age plane, but the overall trends we measure cannot be strongly influcenced.

\subsection{$\dot{M}_{\rm acc}$ vs position}
\begin{figure}
\epsscale{1.1}
\plotone{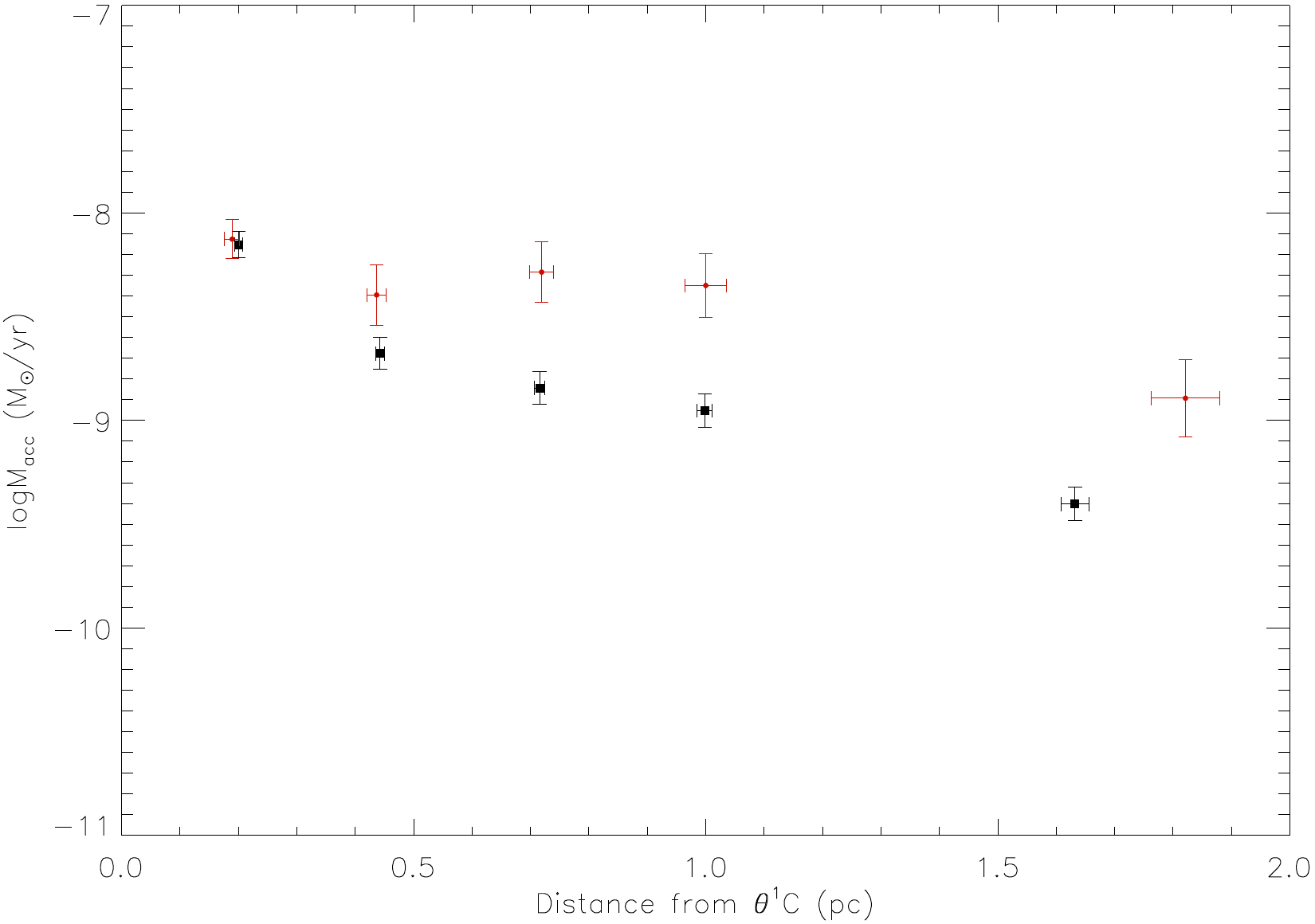}
\caption{Mean values of \macc \ at increasing distances from $\theta^1$ Orionis C. The black points represent the values using all the sources in the sample and dividing them in 5 subsamples of equal number of sources. Red points represent the mean values when selecting for each subsample only the sources with $M_*>0.3 M_\odot$ and 5.5$<\log$(age/yr)$<$7.3. In the first case a trend of lower \macc \ at higher distances is clear, but can be due, at least partially, to selection effects, as we show with the red points, where the distribution is flatter.}
\label{positions}
\end{figure}

We look for spatial variations of $\dot{M}_{\rm acc}$ within the ONC. Such variations, in fact, might imply that the accretion is subject to environmental effects, such as the influence of the massive stars present in the central region of the cluster, in particular $\theta^1$ Orionis C, an O-type star.

In order to do that, we divide our sample in 5 subsamples, each with the same amount of sources, according to the distance of each object from $\theta^1$ C. We show with black squared symbols in Fig.~\ref{positions} the mean values of the \macc \ plotted against the mean value of the distance for each subsample. It is evident that at smaller distances from the central star sources appear to have an higher mean \macc \ that decreases monotonically with the distance. This result, however, does not necessarily indicate a real trend. In fact, around the Trapezium cluster, the detection limit is generally poorer than at larger distances, due to the higher nebular background. Therefore, low-mass stars (whose \macc \ are generally lower, as described in the previous sections) are less easily detected. Therefore, we select in each subsample only the sources with $M_* < 0.3 M_\odot$ and with 5.5 $<\log (t$/yr) $<7.3$, to have similar selection effect in each bin. The result is shown with the red circles in Fig.~\ref{positions}: the values of \macc \  as a function of distance have a much flatter distribution. Only at distances $\gtrsim$ 1.5 pc the mean value of \macc \ is lower. 

%_______________________________________________________
\section{Conclusion}
\label{conclusion}
We have presented a study of mass accretion rates (\macc) in the Orion Nebula Cluster, for an unprecedentedly large ($\sim$ 700 stars) sample of PMS stars. This allowed us to perform a thorough statistical analysis of the dependence of this quantity on the central stellar parameters, and investigate this phase of the stellar mass build-up and disc evolution.

We based our study on HST/WFPC2 photometric data over a large field of view, and derived \macc\ using two different accretion tracers: the ($U-B$) excess and the $L_{H\alpha}$. We study the systematic dependence of \macc\ with respect the age of the sources, their mass and position within the region. In particular, the \macc\ is found to vary with age and mass altogether. Our final relation between all these quantities is given by Equation \ref{fit3D_all} and in Table~\ref{param_fit_table}, and our best fit relation using all the sources in the sample is given by: $\log$(\macc/$M_\odot\cdot$yr)=(-5.12 $\pm$ 0.86) -(0.46 $\pm$ 0.13) $\cdot \log (t$/yr) -(5.75 $\pm$ 1.47)$\cdot \log (M_*/M_\odot)$ + (1.17 $\pm$ 0.23)$\cdot \log (t/$yr$) \cdot \log (M_*/M_\odot)$. We clearly find that the \macc\ increases with stellar mass, and decreases over evolutionary time.

Interestingly, we also find evidence that for more massive stars the decay of \macc\ with time is much slower than for lower stellar masses. Similarly, for older stars, the dependence of \macc\ with $M_*$ appears significantly steeper. This might imply that these objects are not in the asymptotic regime (i.e. when $t\gg t_\nu$, where $\dot{M}_{acc}\propto t^{-\eta}$), or that the hypothesis of a simple dependence of the viscosity $\nu\propto R^\gamma$ is probably not compatible with our observations. In particular, we suggest significant discrepancies of our results with respect to the self-similar parametrization for sources with masses higher than $\sim 0.5 M_\odot$.

%_______________________________________________________
\acknowledgments
This work was made possible by GO program 10246 of the {\it Hubble Space Telescope}, which is operated by the Space Telescope Science Institute.

C.F.M. would like to thank James Pringle and Philip J. Armitage for insightful discussions.

{\it Facilities}: HST (WFPC2)

\bibliographystyle{apj}
\bibliography{ms}

\clearpage

%
%_______________________________________________________
\appendix

%
%_______________________________________________________
\section{Red Leak}
\label{red_leak}

\begin{figure}
\epsscale{.50}
\plotone{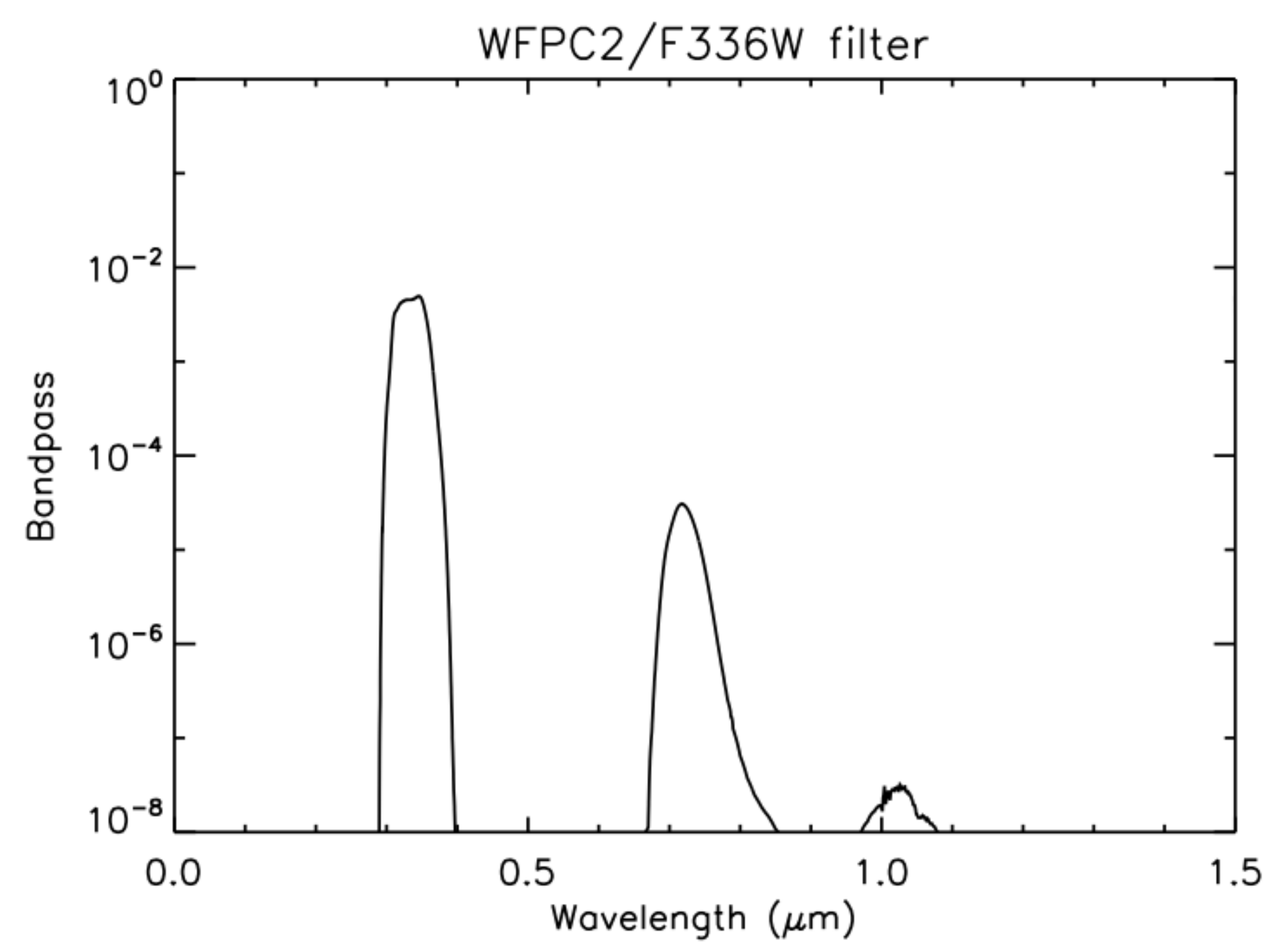}
\caption[Bandpass of the filter F336W/WFPC2]{Bandpass of the filter F336W/WFPC2. Note the red leak component at $\sim0.75\mu$m, where the filter has a window of transmission with 1\% of the peak transmission at $\sim 0.35 \mu$m. }
\label{leak}
\end{figure}

As mentioned in Sec.~\ref{observations}, the filter F336W of WFPC2 is affected by red leak. This is evident from Fig.~\ref{leak}, where the band profile is shown.
At $\lambda\sim$7300~\AA, the filter shows a transmission window whose peak throughput is $\sim$0.5\% that in the $U$-band wavelength range.
There is also a third peak at $\sim$ 1 $\mu$m, but this is completely negligible, since its bandpass is less than 10$^{-5}$ of the main peak and the detector Quantum Efficiency is nearly zero.

The presence of red leak in F336W does not significantly affect the stellar photometry of blue sources, as most of the  flux is emitted at short wavelengths. For red objects, however, the leak contribution may be significant and, in extreme cases, even dominant. In order to accurately account for the $U$-band fluxes of our ONC sources, and prevent introducing a systematic overestimate of the accretion luminosity (\lacc) and mass accretion rate (\macc), we must estimate and correct for the spurious red-leak excess.

The number of photons collected in the leaking part of the filter depends solely on the observed flux at the wavelength of the leak (referred to as $U_{\rm leak}$, in magnitudes). To estimate this flux, we considered our WFPC2 $I$-band photometry and computed the color term ($U_{\rm leak} - I$) by means of synthetic photometry. 
As this color term depends mainly on $T_{\rm eff}$ and $A_V$, we computed it for each source considering the stellar parameters from \citet{DaRio12}, and assuming the atmosphere models from \citet{BT-Settle}, corrected as explained in Appendix~\ref{app_model_calibration}.

We then converted $U_{\rm leak}$ from magnitudes to photons/second. To this end, one must derive the zero-point of the leak window of the filter, i.e, the number of photons passing through the leak for a magnitude m=0. This zero point is simply the zero-point of the whole F336W filter (ZP$_{F336W}$, \citealt{CTEcorr}), scaled to a term equal to the fraction of $U$-band photons that pass through the leak.
We derived this latter term from the calibrated spectrum of Vega of \citet{bohlin2007} and the measured F336W throughput. We find that the fraction of photons in the red leak part of the spectrum of Vega is 0.551\% of the total number of $U-$band photons. Therefore, we obtain $counts_{\rm leak} = 10^{-0.4(U_{\rm leak} - ZP)}\cdot 0.00551.$

Finally, we considered our observed $U$ magnitudes, converted them in units of counts (using again ZP$_{F336W}$), subtracted the leak contribution, and transformed them once again in magnitudes. The result is the leak-corrected $U-$band magnitude, or $U_{\rm noleak}$. The leak offset in magnitude, for our stellar sample, has a mean value of about 0.5 mag. We refer in all the text to the $U$ band as the corrected one, using the extended notation $U_{\rm noleak}$ only when needed, for clarity.

%
%_______________________________________________________
\section{Isochrone calibration}
\label{app_model_calibration}
In order to calibrate the appropriate isochrone for our sample, we start by considering a family of synthetic spectra, the BT-SETTL from \citet{BT-Settle}. As shown in their work, these models seem to reproduce the observed colors of stars and brown dwarfs in the $J$ and $K$ band. However, their validity in the optical range has not been demonstrated. We therefore test their accuracy taking advantage of the $J$ band photometry on the ONC from \citet{ISPI}.
The concept is the following: assuming the validity of the predicted $J$-band colors as a function of $T_{\rm eff}$, if the predicted $I$-band colors is also correct, then the synthetic $(I-J)$ should also agree with the data. Otherwise, a correction to the synthetic $I$-band magnitude is required. Subsequently, the same procedure can iterated to the $B$ and $U$ bands as well.

\begin{figure}
\epsscale{0.5}
\plotone{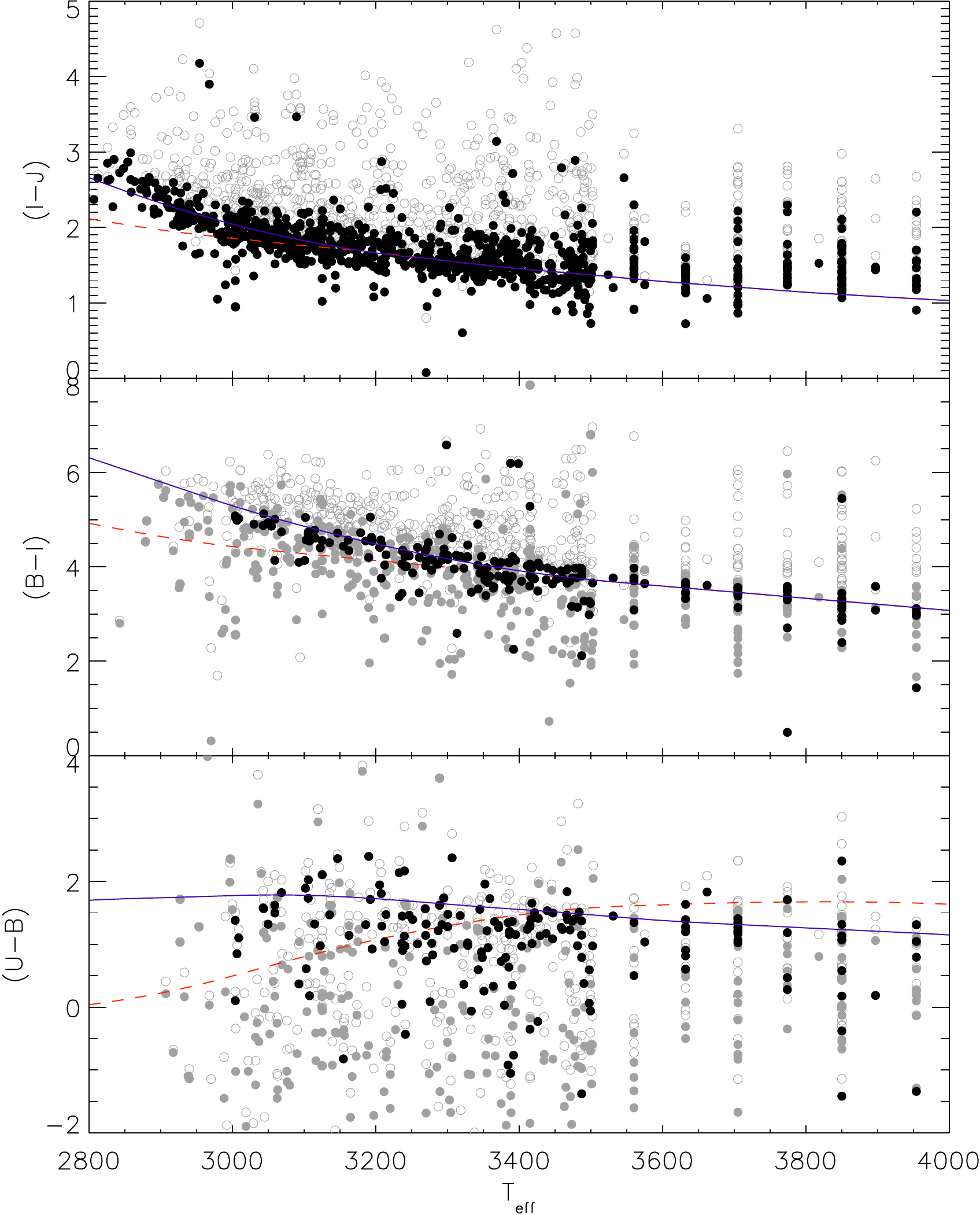}
\caption{Procedure used to calibrate the models by \citet{BT-Settle} in our photometric bands. The red dashed line represents the color computed trough synthetic photometry on the BT-SETTL models. Grey open circles are the observed colors of the sources, grey filled one the corrected values according to $A_V$ from \citet{DaRio12} and black filled points are corrected values of low accretors according to \citet{DaRio}. We use the latter ones to determine the empirical calibration, that is the blue line.}
\label{calibration}
\end{figure}

Fig.~\ref{calibration} illustrates this test: the upper panel shows the observed (grey open circles) and extinction-corrected (black dots) photometry of our ONC members, assuming the $A_V$ from \citet{DaRio12}, and limited to the candidate weak accretors \citep{DaRio,DaRioI}, i.e. sources with $\log (L_{\rm acc}/L_{\rm tot})_{DaRio10} <$ -1.5 and EW$_{H\alpha,DaRio09}<$ 10 \AA. Under these assumptions, the black dots, therefore, trace the photospheric colors of ONC sources of different temperatures.
The red dashed line represents a 1~Myr synthetic isochrone from the models of \citet{Baraffe}, converted into colors assuming the BT-SETTL atmosphere models models. We find that the predicted colors are systematically underestimated for $T_{\rm eff}\lesssim3200$~K.
We thus interpolate the empirical $I-J$ color locus (blue solid line), and determine the offset between the two. We calibrated the $I$-band magnitudes of the isochrone as a function of $T_{\rm eff}$, by applying this correction to the synthetic $I-$band magnitudes.

From the calibrated $I$-band photospheric magnitudes, we iterate the same procedure first to the $B-I$ colors, in order to calibrate $B$ (Fig.~\ref{calibration}, middle panel), then to the $U-B$ color, to calibrate the intrinsic $U$ magnitude (lower panel). The selection of the candidate non-accretors is particulary critical for these two bands, which are highly influenced by accretion excesses.

We stress that for practical reasons we performed a calibration of the synthetic isochrone in terms of magnitudes, but for our purposes (the analysis of our 2CD) only the accuracy of the colors as a function of temperature are relevant. Also, colors are independent on distance or small differences in stellar radii, therefore a luminosity spread for ONC members, due to e.g., a spread in stellar ages, does not affect our method.
We recap the empirical estimate of the WFPC2 colors found for the calibration of the BT-SETTL models in Table \ref{corrections}.

\end{document}